\documentclass{LMCS}

\usepackage{amsmath}
\usepackage{amssymb}
\usepackage{mathrsfs}
\usepackage{bussproofs}
\usepackage{enumerate,hyperref}

\newcommand{\Long}  [2]                {#2}
\newcommand{\PaperState}               {\Long}
\newcommand{\IfPaperState}  [2]        {\PaperState{#1}{#2}}

\newcommand{\EM}                       { {\tt EM} }
\newcommand{\HA}                       { {\tt HA} }
\newcommand{\Nat}                      { {\tt N} }
\newcommand{\Bool}                     { {\tt Bool} }
\newcommand{\State}                    { {\tt S} }
\newcommand{\NatSet}                   {\mathbb{N}}
\newcommand{\BoolSet}                  {\mathbb{B}}
\newcommand{\StateSet}                 {\mathbb{S}}
\newcommand{\SystemT}                  {\mathcal{T}}
\newcommand{\Class}                    {\mbox{\tiny Class}}
\newcommand{\Learn}                    {\mbox{\tiny Learn}}
\newcommand{\SystemTState}             {\SystemT_\State}
\newcommand{\SystemTClass}             {{\SystemT_{\Class}}}
\newcommand{\SystemTLearn}             {{\SystemT_{\Learn}}}
\newcommand{\Language}                 {\mathcal{L}}
\newcommand{\LanguageClass}            {\Language_{\Class}}
\newcommand{\LanguageLearn}            {\Language_{\Learn}}
\newcommand{\Realizer}                 {{\mbox{\tiny Real}}}
\newcommand{\StructureN}               {{\mathcal N}}
\newcommand{\Add}                      { {\mbox{Add}} }
\newcommand{\add}                      { {\mbox{add}} }
\newcommand{\fix}                      { {f} }
\newcommand{\proj}                     { {p} }
\newcommand{\substitution} [1]         { {\overline{#1}} }
\newcommand{\True}                     { {\tt{True}} }
\newcommand{\False}                    { {\tt{False}} }
\newcommand{\makenum}        [1]       { {\underline{#1}} }
\newcommand{\makestate}      [1]       { {\underline{#1}} }
\newcommand{\Succ}                     { {\mbox{S}} }
\newcommand{\CupSem}                   { {\mathcal U} }

\def\doi{6 (3:19) 2010}
\lmcsheading%
{\doi}
{1--22}
{}
{}
{Nov.~15, 2009}
{Sep.~\phantom07, 2010}
{}

\begin{document}

\title{Interactive Learning-Based Realizability
for Heyting Arithmetic with
$\EM_1$ \IfPaperState{(extended abstract)}{}}

\author[F.~Aschieri]{Federico Aschieri\rsuper a}
\address{{\lsuper a}C.S.~Department, Universit\`a di Torino, Italy, and\hfill\break
  School of EECS, Queen Mary, University of London, UK}
\email{federico.aschieri@di.unito.it}
\author[S.~Berardi]{Stefano Berardi\rsuper b}
\address{{\lsuper b}C.S.~Department, Universit\`a di Torino, Italy}
\email{berardi@di.unito.it}

\begin{abstract}
We apply to the semantics of Arithmetic the idea of ``finite
approximation'' used to provide computational interpretations of
Herbrand's Theorem, and we interpret classical proofs as constructive
proofs (with constructive rules for $\vee, \exists$) over a suitable
structure $\StructureN$ for the language of natural numbers and maps
of G\"odel's system $\SystemT$. We introduce a new Realizability
semantics we call ``Interactive learning-based Realizability'', for
Heyting Arithmetic plus $\EM_1$ (Excluded middle axiom restricted to
$\Sigma^0_1$ formulas). Individuals of $\StructureN$ evolve with time,
and realizers may ``interact'' with them, by influencing their
evolution. We build our semantics over Avigad's fixed point result,
but the same semantics may be defined over different constructive
interpretations of classical arithmetic (Berardi and de' Liguoro use
continuations). Our notion of realizability extends intuitionistic
realizability and differs from it only in the atomic case: we
interpret atomic realizers as ``learning agents''.
\end{abstract}


\maketitle

\section{Introduction}
From now on, we will call $\HA$ Heyting Intuitionistic Arithmetic, with a language including one symbol for each primitive recursive predicate or function. We call $\Sigma^0_1$-formulas the set of all formulas $\exists x.P(x,y)$ for some primitive recursive predicate $P$, and $\EM_1$ the {\em Excluded middle axiom restricted to $\Sigma^0_1$-formulas}. For a detailed study of the intuitionistic consequences of the sub-classical axiom $\EM_1$ we refer to \cite{BerCoqKohlLICS}.

In this paper we give the full version of Aschieri and Berardi \cite{AB} and we extend Berardi and de' Liguoro (\cite{Ber2005}, \cite{Berardi}) notion of atomic realizability - originally conceived for quantifier free primitive recursive Arithmetic plus $\EM_1$ - to full predicate logic, namely Heyting Arithmetic with $\EM_1$ ($\HA + \EM_1$). Our idea is to interpret classical proofs as constructive proofs on a
suitable structure $\StructureN$ for natural numbers and maps of G\"odel's system $\SystemT$, by applying to the semantics of Arithmetic the idea of ``finite approximation'' used to interpret Herbrand's Theorem. We extend intuitionistic realizability to a new notion of realizability, which we call ``Interactive learning-based Realizability". We provide a term assignment for the standard natural deduction system of $\HA + \EM_1$, which is surprisingly equal in all respects to that of $\HA$, but for the fact that we have non-trivial realizers for
atomic formulas and a new realizer for $\EM_1$.

Our semantics is ``local'': we do not introduce a global variable representing the goal, as in continuation interpretation, in Friedman's $A$-translation and in Krivine's Classical Realizability. We interpret classical proofs ``locally'' and step-by-step, in order to solve a major problem of all computational interpretations: global illegibility, which means that, even for simple classical proofs, it is extremely difficult to understand how each step of the extracted program is related to the ideas of the proof, and what it is the particular task performed by each subprogram of the extracted program. The main sources of inspiration of this paper are works of
Kleene, Hilbert, Coquand, Hayashi, Berardi and de' Liguoro and Avigad.

\textit{Intuitionistic Realizability revisited.} In \cite{Kleene}, Kleene
introduced the notion of realizability, a formal semantics for
intuitionistic arithmetic. Later, Kreisel \cite{Kreisel} defined modified realizability, the same notion but with respect to a typed lambda calculus instead of Kleene's formalism of partial recursive functions.  Realizability is nothing but a formal
version of Heyting semantics for intuitionistic logic, translated
into the language of arithmetic.

Intuitively, realizing a closed arithmetical formula $A$  means
exhibiting a computer program - called realizer - able to calculate
all  relevant information about the truth of $A$.  Hence, realizing
a formula $A\lor B$ means realizing $A$ or  realizing $B$, after
calculating \textit{which one} of the two is actually realized;
realizing a formula $\exists x A(x)$ means computing a numeral $n$ - called a witness -  and realizing $A(n)$.

These two cases are indeed the only ones in which we have relevant
information to calculate about the truth of the corresponding
formula, and there is a  decision to be made: realizing a formula
$\forall x A$ means exhibiting an algorithm which takes as input a numeral $n$ and gives as output realizers of $A(n)$;
realizing a formula $A\wedge B$ means realizing $A$ and realizing $B$; realizing $A\rightarrow B$ means providing an algorithm which takes as input realizers of $A$ and gives realizers of $B$; in these cases we provide no information about the formula we realize and we only take the inputs we will use for realizing existential or disjunctive formulas. Finally, realizing an atomic formula means that the formula is true: in this case, the realizer does nothing at all.

\IfPaperState{}{Hence, intuitionistic realizability closely follows
Tarski's definition of truth - the only difference being
effectiveness: for instance, while Tarski, to assert that $\exists
x
A$ is true, contented himself to know that there exists some $n$
such that  $A(n)$ is true, Kleene asked for a program that
calculates an $n$ such that $A(n)$ is
true.}

Intuitionistic natural deduction rules are perfectly suited to
preserve realizability. In order to actually build realizers from
intuitionistic natural deductions, it suffices to give realizers for
the axioms. Since our goal is to interpret classical connectives
using Heyting and Kleene interpretation of intuitionistic
connectives, then a first, quite naive idea would be the following:
if we devised realizers for Excluded Middle, we would be able to extend
realizability to all classical arithmetic.

Unfortunately, from the work of Turing it is
well known that not every instance of Excluded Middle is
realizable.
If $Txyz$ is Kleene's predicate, realizing $\forall x \forall y.
\exists z Txyz \lor \forall z \neg Txyz$ implies exhibiting an
algorithm which for every $n,m$ calculates whether or not the
$n$-th
Turing machine halts on input $m$: the halting problem would be
decidable. Hence, there is no hope of computing with effective
programs all the information about the truth  of Excluded
Middle.

However, not all is lost. A key observation is
the
following. Suppose we had a realizer
$O$ of the Excluded Middle and we made a natural deduction of a
formula $\exists x A$ actually using  Excluded Middle; then, we
would be able to extract from the proof a program $u$, containing
$O$ as subprogram, able to compute the witness for $\exists x A$.
Given the effectiveness of $u$, after a finite number of steps -
and
more importantly, after a finite number of calls to $O$ - $u$ would
yield the required witness. It is thus clear that $u$, to perform
the calculation, would use only a \textit{finite piece of
information about the Excluded Middle}. This fundamental fact gives us hope: maybe there is not always necessity of fully realizing Excluded Middle, since in finite computations only a finite amount of information is used. If we were able to gain that finite information during the computation, as it is the case in the proof of Herbrand's Theorem, we could adapt intuitionistic realizability to Classical Logic.

\textit{Herbrand's Theorem and the idea of ``finite approximation".} (A corollary of) Herbrand's Theorem says that if a universal first order theory $T$, in a suitable language supporting definition by cases, proves a statement $\exists x P(x)$, then one can extract from any proof a term $t$ and closed instances $A_1,\ldots, A_n$ of some universal formulas of $T$ such that $A_1\land \ldots \land A_n\rightarrow P(t)$ is a propositional tautology. So, even using classical logic, one can define witnesses. The problem is that the functions occurring in $t$ may not be computable, because the language of $T$ is allowed to contain arbitrary functions. However, given the finiteness of the information needed about any function used during any finite computation of $t$, in order to carry out actual calculations one would only have to find finite approximations of the non-computable functions involved, thus recovering effectiveness. We choose to follow this intuition:  we will add non-computable functions to our language for realizers and exploit the existence of these ideal objects in order to find concrete computational solutions.

This general idea dates back to Hilbert's $\epsilon$-substitution method (for a neat reformulation of the $\epsilon$-method see for example Avigad \cite{Avigad}). As noted by Ackermann \cite{Ackermann}, the $\epsilon$- substitution method may be used to compute witnesses of provable existential statements of first order Peano Arithmetic. The procedure is simple: introduce Skolem functions (equivalently, $\epsilon$-terms) and correspondent quantifier free Skolem axioms in order to reduce any axiom to a quantifier free form; take a $PA$-proof of a sentence $\exists x P(x)$ and translate it into a proof using as axioms only universal formulas; then apply Herbrand's theorem to the resulting proof, obtaining a quantifier free proof of $P(t)$, for some term $t$ of the extended language; finally, calculate a suitable finite approximation of the Skolem functions occurring in $t$ and calculate from $t$ an $n$ such that $P(n)$ holds.

However, while proofs in quantifier free style are very simple combinatorial objects, they lose the intuitive appeal, the general concepts, the structure of high level proofs. Hence, it may be an impossible task to understand extracted programs. Moreover we have a computational syntactic method but no semantics of proofs and logical operators based on the idea of ``finite approximation'', as the realizability interpretations are based on the idea of ``construction''. However, in the $\epsilon$-method, albeit only for quantifier free formulas, we see in action the method of intelligent \textit{learning}, driven by the Skolem axioms used in the proofs. One of the aims of this paper is to extend this ``semantics of learning" from atomic propositions to individuals, maps, logical connectives and quantifiers of full natural deduction proofs. An important contribution comes from Coquand \cite{Coquand}.

\textit{Coquand's Game Semantics for Classical Arithmetic.} Computing all relevant information about the truth of a
given formula $A$ is not always possible. In \cite{Coquand} and in the context of game semantics, Coquand introduced a new key idea around this problem: the correspondence between backtracking and ``learning'', a refinement of the idea of ``finite approximation''. If we cannot
compute all the right information about the truth of a formula, maybe we could do this if we were allowed to make finitely many mistakes and to learn from them.

Suppose, for instance, we have the formula $\forall x. \exists y
Pxy\lor \forall y \neg Pxy$, but we have no algorithm which, for all numeral $n$ given as input, outputs false if $\forall y \neg Pny$ holds and outputs true if $\exists y Pny$ holds. Then we may describe a learning algorithm $r$ as follows.  Initially, for all $n$ given as input, $r$ outputs false. Intuitively, $r$ is initially persuaded - following the principle ``if I don't see, I do not believe" - that for all numeral $n$ there is no numeral $m$ such that $Pnm$ holds. Hence, when asked for his opinion about the formula $\exists y Pny\lor \forall y \neg Pny$, $r$ always says: $\exists y Pny$ is false. However, if someone - an opponent of $r$ - to show that $r$ is wrong, comes out with an $m$ such that $Pnm$ holds, $r$ realizes indeed to be mistaken, and stores the information ``$Pnm$ is true". Then, the next time being asked for an opinion about $\exists y Pny\lor \forall y \neg Pny$, $r$ will say: true. In other words, such $r$, after at most one ``mind changing'', would be able to learn the correct answer to any question of the form: ``which one among $\exists y Pny$, $\forall y \neg Pny$ does hold?". This is actually learning by counterexamples and is the key idea behind
Coquand's semantics.

Our question is now: can we formulate a realizability notion based on learning by counterexamples in order to extend Kreisel's interpretation to all individuals, maps and connectives of the sub-classical Arithmetic $\HA +  \EM_1$? Following Hayashi \cite{Hayashi}, in our solution we modify the notion of individual, in such a way that individuals change with time, and realizers ``interact'' with them.



\textit{Hayashi's Proof Animation and Realizability.}
\IfPaperState{}{
}
In \cite{Hayashi}, Hayashi explains a notion of realizability for a sub-classical arithmetic, called limit computable mathematics.
Basing his analysis on ideas of Gold \cite{Gold}, he defines a Kleene's style
notion of realizability equal to the original one but for the fact
that the notion of individual changes: the
witnesses of existential and disjunctive formulas are calculated by
a stream of guesses and ``learned in the limit'' (in the sense that
the limit of the stream is a correct witness). An individual $a$
is therefore a computable map $a:\mathbb{N} \rightarrow \mathbb{N}$,
with $a(t)$ representing the value of the individual at time $t$.

For instance, how would Hayashi realize the formula $\forall x. \exists y Pxy\lor \forall y \neg Pxy$? He would define an algorithm $H$ as follows. Given any numeral $n$, $H$ would calculate the truth value of $\forall y\leq n Pny$. Then the correct answer to the question: ``which one among $\exists y Pny$, $\forall y \neg Pny$ does hold?" is learned in the limit by computing $P(n,0)$, $P(n,1)$, $P(n,2)$,\ldots, $P(n,k)$,\ldots and thus  producing a stream of guesses either of the form false, false, false,\ldots, true, true,\ldots, true,\ldots or of the form false, false, false, \ldots, false, \ldots, the first stabilizing in the limit to true, the second to false. Hayashi's idea is to perform a completely
blind
and exhaustive search: in such a way, the correct answer is
guaranteed to be eventually learned (classically). Hayashi's realizers do not
learn in an efficient way: in Hayashi's notion of realizability the
only learning device is to look through all possible cases.
Instead,
we want to combine the idea of individual as limit, taken from Hayashi, with notion of learning in which the stream of guesses is
{\em driven by the proof itself}, as in Coquand's game semantics. For the quantifier-free fragment, this was done by Berardi \cite{Ber2005} and Berardi-de' Liguoro \cite{Berardi}.

\textit{Realizability Based on Learning: Berardi-de'
Liguoro interpretation.}
We explain the paper \cite{Berardi} using Popper's ideas \cite{Popper} as a metaphor.
According to Popper, a scientific theory relies on a set of
unproved - and unprovable - hypotheses and, through logic, makes
predictions suitable to be falsified by experiments. If a prediction is falsified,
some hypothesis is
incorrect. In front of a counterexample to a theory's prediction,
one must modify the set of hypotheses and  build a better theory,
which will be tested by experiments, and so on. Laws of Nature are
universal statements, that cannot be verified, but are suitable to
falsification.
We may explain the link between falsifiable hypotheses and
$\EM_1$. For every $n$, given an instance $\exists y.Pny  \vee
\forall
y.\neg Pny$  of $\EM_1$ (with $P$ atomic), we may formulate an
hypothesis about which side of the disjunction is true. If we know
that $Pnm$ is true for some $m$, we know that $\exists y.Pny$ is
true. Otherwise we may assume $\forall y. \neg Pny$ as hypothesis,
because it is a falsifiable hypothesis.


We formalize the process of making hypotheses about
$\EM_1$ by a finite state of knowledge, called $S$,
collecting the instances $Pnm$ which we know to hold, e.g. by direct calculation.  If we have evidence that $Pnm$ holds for some $m$ (that is, $Pnm\in S$) we know that $\exists y Pny$ is true; in the other case, we assume that $\forall y \neg Pny$ is true. So $S$ defines a set of hypotheses on $\EM_1$, of the form $\forall y \neg Pny$: universal falsifiable statements. Using $S$ a realizer $r$ may effectively decide which side of a given instance of $\EM_1$ is true,
at the price of making mistakes: to decide if $\forall y \neg Pny$ is true, $r$ looks for any $Pnm$ in the finite state $S$ and outputs ``false'' if the research is successful, ``true'' otherwise. If and when from an hypothesis $\forall y \neg Pny$ we obtain some false conclusion $\neg Pnm$, the realizer $r$ returns the additional knowledge: ``$Pnm$ is true'', to be added to $S$.

%
%


\textit{Extending Berardi-de'
Liguoro interpretation to $\HA+\EM_1$.} In our paper, we interpret each classical proof $p$ of $A$ in $\HA + \EM_1$ by a ``learning realizer'' $r$. $r$ returns a ``prediction'' of the truth of this formula, based on the information in $S$, and some additional knowledge in the case the prediction is effectively falsified.  For example, in front of
a formula $\exists x. A \wedge B$, a realizer $r$ predicts that $A(n)\wedge B(n)$ is true for some numeral $n$ (and since $n$ depends on $s$, in our model we change the notion of individual, interpreting ``numbers'' as  computable maps from the set of bases of knowledge to $\NatSet$).
Then $r$ predicts, say, that $B(n)$ is true, and so on, until $r$ arrives at some atomic formula, say $\neg Pnm$. Either $Pnm$ is actually true, or $r$ is able to effectively find one or more flawed hypothesis $\forall x.\neg Q_1n_1x, \ldots, \forall x.\neg Q_kn_kx$ among the hypotheses used to predict that $Pnm$ is true, and for each flawed hypothesis one counterexample $ Q_1n_1m_1,\ldots, Q_kn_km_k$. In this case, $r$ requires to enlarge our state of knowledge $S$ by including the information ``$Q_1n_1m_1$ is true", \ldots, ``$Q_kn_km_k$ is true".

Our Interactive Realizability differs from Intuitionistic
Realizability in the notion of individual (the value of an individual may depend on our knowledge state), and in the
realizability relation for the atomic case. In our interpretation, to realize an atomic formula does not mean that the formula is true, but that the realizer requires to extend our state of knowledge $S$ if the formula is not true. The realizer is thought as a learning device. Each extension of $S$ may change the value of the individuals which are parameters of the atomic formula, and therefore may make the atomic formula false again. Then the realizer requires to extend $S$ again, and so forth. The convergence of this ``interaction'' between a realizer and a group of individuals follows by Avigad's fixed point thm. \cite{Avigad} (a constructive proof may be found in \cite{Ber2005}), and it is the analogue of the termination of Hilbert's $\epsilon$-substitution method.



\textit{Why the Arithmetic $\HA + \EM_1$} instead of considering the full Peano Arithmetic? We have two main reasons. First, we observe that $\EM_1$ enjoys a very good property: the information about its truth can be computed in the limit, in the sense of Gold \cite{Gold}, as we saw en passant when discussing Hayashi's realizability. This implies that witnesses for existential and disjunctive statements too can be  learned in the limit, as shown in Hayashi \cite{Hayashi}. In a forthcoming paper we show that realizers which we will be able to extract from proofs have a straightforward interpretation as winning strategies in 1-Backtracking games \cite{BerCoq}, which are the most natural and simple instances of Coquand's style games. Secondly, a great deal of mathematical theorems are proved by using $\EM_1$ alone (\cite{BerCoqKohlLICS}, \cite{BerAPAL}).


\textit{Plan of the Paper}. The paper is organized as follows. In \S \ref{section-TheTermCalculus} we define the term calculus in which our realizers will be written: a version of G\"odel's system $\SystemT$, extended with some syntactic sugar, in order to represent bases of knowledge (which we shall call states) and to manipulate them. Then we prove a convergence property for this calculus (as in Avigad \cite{Avigad} or in \cite{Ber2005}). In \S \ref{section-ALearningBasedRealizability}, we introduce the notion of realizability and prove our Main Theorem, the Adequacy Theorem: ``if a closed arithmetical formula is provable in $\HA + \EM_1$, then it is realizable''.
In \S \ref{section-conclusion} we conclude the discussion about our notion of realizability by comparing it with other notions of realizability for classical logic, then we consider some possible future work.


\section{The Term Calculus}\label{section-TheTermCalculus}

In this section we formalize the intuition of ``learning realizer'' we discussed in the introduction.

We associate to any instance $\exists y Pxy\lor \forall y \neg Pxy$ of $\EM_1$ (Excluded Middle restricted to $\Sigma^0_1$-formulas) two functions $\chi_P$ and $\varphi_P$. The function $\chi_P$ takes a knowledge state $S$, a numeral $n$, and it returns a guess for the truth value of $\exists y. Pny$. When this guess is ``true'' the function $\varphi_P$ returns a witness $m$ of $\exists y.Pny$. The guess for the truth value of $\exists y. Pny$ is computed w.r.t. the knowledge state $ S $, and it may be wrong. For each constant $s$ denoting some knowledge state $ S $, the function $\lambda x:\Nat.\chi_P( s ,x)$ is some ``approximation'' of an ideal map $\lambda x:\Nat.X_P(x)$, the \emph{oracle} returning the truth value of $\exists y. Pxy$. In the same way, the function $\lambda x:\Nat.\phi_P( s ,x)$ is some ``approximation'' of an ideal map $\lambda x:\Nat.\Phi_P(x)$, the \emph{Skolem map} for $\exists y. Pxy$, returning some $y$ such that $Pxy$ if any, and $0$ otherwise. The Skolem axioms effectively used by a given proof take the place of a set of experiments testing the correctness of the predictions made by $\varphi_P(s,x), \chi_P(s,x)$ about $X_P(x),\Phi_P(x)$ (we do not check the correctness of $\varphi_P, \chi_P$ in an exhaustive way, but only on the values required by the Skolem axioms used by a proof).

Our Term Calculus is an extension of G\"odel's system $\SystemT$. For a complete definition of $\SystemT$ we refer to Girard \cite{Girard}. $\SystemT$ is simply typed $\lambda$-calculus, with atomic types $\Nat$ (representing the set $\NatSet$ of natural numbers) and $\Bool$ (representing the set $\BoolSet = \{\mbox{True},\mbox{False}\}$ of booleans), product types $T \times U$ and arrows types $T \rightarrow U$, and pairs $\langle.,.\rangle$, projections $\pi_0, \pi_1$, conditional ${\tt if}_T$ and primitive recursion $R_T$ in all types, and the usual reduction rules $(\beta),(\pi),(if),(R)$ for $\lambda$, $\langle .,.\rangle,{\tt if}_T,R_T$. From now on, if $t, u$ are terms of $\SystemT$ with $t=u$ we denote provable equality in $\SystemT$. If $k \in \NatSet$, the numeral denoting $k$ is the closed normal term $\makenum{k} = \Succ^k(0)$ of type $\Nat$. We denote numerals in $\SystemT$ by $n,m$, and natural numbers with $i,j,k,h, \ldots \in \NatSet$. All closed normal terms of type $\Nat$ are a numeral. We denote with ${\True},{\False}:\Bool$ the boolean constants of $\SystemT$. Any closed normal term of type $\Bool$ in $\SystemT$ is ${\True}$ or ${\False}$.



We introduce a notation for ternary projections: if $T = A \times (B \times C)$, with $p_0, p_1, p_2$ we respectively denote the terms $\pi_0$, $\lambda x:T.\pi_0(\pi_1(x))$, $\lambda x:T.\pi_1(\pi_1(x))$.
If $u = \langle u_0,\langle u_1,u_2\rangle \rangle : T$, then $\proj_iu=u_i$ in $\SystemT$ for $i=0,1,2$. We abbreviate $\langle u_0,\langle u_1,u_2\rangle \rangle :T $ with $\langle u_0,u_1,u_2\rangle : T$. We formalize the idea of ``finite information about $\EM_1$'' by the notion of {\em state of knowledge}.\vfill\eject

\begin{defi}[States of
Knowledge and Consistent Union]\label{definition-StateOfKnowledge}\hfill
\begin{enumerate}[(1)]
\item
A $k$-ary {\em predicate} of $\SystemT$ is any closed normal term $P:\Nat^{k}\rightarrow \Bool$ of $\SystemT$.

\item
An atom is any triple $\langle P,\vec{n},{m}\rangle $, where $P$ is a $(k+1)$-ary predicate, and $\vec{n},m$ are $(k+1)$ numerals, and $P\vec{n}m = \True$ in $\SystemT$.

\item
Two atoms $\langle P,\vec{n},{m}\rangle $, $\langle P',\vec{n'},{m'}\rangle $ are {\em consistent} if $P = P'$ and $\vec{n} = \vec{n'}$ in $\SystemT$ imply $m = m'$.

\item
A state of knowledge, shortly a {\em state}, is any finite set $S$ of pairwise consistent atoms.

\item
Two states $S_1, S_2$ are consistent if $S_1 \cup S_2$ is a state.

\item
$\StateSet$ is the set of all states of knowledge.
\item
The {\em consistent union} $S_1 \CupSem S_2$ of $S_1, S_2 \in \StateSet$ is $S_1 \cup S_2 \in \StateSet$ minus all atoms of $S_2$ which are inconsistent with some atom of $S_1$.
\end{enumerate}
\end{defi}

\noindent We think of an atom $\langle P,\vec{n},{m}\rangle $ as the code of a witness for $\exists y.P(\vec{n},y)$. Consistency condition allows at most one witness for each $\exists y.P(\vec{n},y)$ in each knowledge state $S$. Two states $S_1, S_2$ are consistent if and only if each atom of $S_1$ is consistent with each atom of $S_2$.

$S_1 \CupSem S_2$ is an non-commutative operation: whenever an atom of $S_1$ and an atom of $S_2$ are inconsistent, we arbitrarily keep the atom of $S_1$ and we reject the atom of $S_2$, therefore for some $S_1, S_2$ we have $S_1 \CupSem S_2 \not = S_2 \CupSem S_1$. $\CupSem$ is a ``learning strategy'', a way of selecting a consistent subset of $S_1 \cup S_2$. It is immediate to show that $\CupSem$ is an associative operation on the set of consistent states, with neutral element $\emptyset$, with upper bound $S_1 \cup S_2$, and returning a non-empty state whenever $S_1 \cup S_2$ is non-empty.

\begin{lem} \label{lemma-Cup}
Assume $i \in \NatSet$ and $S_1, \ldots, S_i \in \StateSet$.
\begin{enumerate}[\em(1)]
\item
$S_1 \CupSem \ldots \CupSem S_i \subseteq S_1 \cup \ldots \cup S_i$

\item
$S_1 \CupSem \ldots \CupSem S_i = \emptyset$ implies $S_1 = \ldots = S_i = \emptyset$.
\end{enumerate}
\end{lem}


In fact, the whole realizability Semantics is a Monad \cite{BerardiLiguoroMonadi}. In \cite{BerardiLiguoroMonadi}, it is proved that our realizability Semantics is parametric with respect to the definition we choose for $\CupSem$. Any associative operation $\CupSem$, with neutral element $\makestate{\emptyset}$ and satisfying the two properties of Lemma \ref{lemma-Cup}, defines a different but sound realizability Semantics, corresponding to a different ``learning strategy''. An immediate consequence of Lemma \ref{lemma-Cup} is:

\begin{lem} \label{lemma-consistency}
Assume $S, S_1, S_2 \in \StateSet$.
\begin{enumerate}[\em(1)]
\item
If $S$ is consistent with $S_1, S_2$, then $S$ is consistent with $S_1 \CupSem S_2$.
\item
If $S$ is disjoint with $S_1, S_2$, then $S$ is disjoint with $S_1 \CupSem S_2$.
\end{enumerate}
\end{lem}


For each state of knowledge $S$ we assume having a unique constant $s = \makestate{S}$ denoting it: for instance, $\makestate{\emptyset}$ is a state constant denoting the empty state. We define with $\SystemTState  = \SystemT + \State + \{\makestate{S}|S \in \StateSet\}$ the extension of $\SystemT$ with one atomic type $\State$ denoting $\StateSet$, and a constant $s = \makestate{S} : \State$ for each $S \in \StateSet$, and {\em no} new reduction rule. We denote states by $S, S', \ldots$ and state constants by $s, s', \ldots$. Any closed normal form of type $\Nat, \Bool, \State$ in $\SystemTState$ is, respectively, some numeral $n$, some boolean $\True, \False$, some state constant $s$. Computation on states will be defined by some suitable set of algebraic reduction rules we call ``functional''.

\begin{defi}\label{definition-functional} (Functional set of rules)
Let $C$ be any set of constants, each one of some type $A_1\rightarrow \ldots \rightarrow A_n\rightarrow A$, for some $A_1,\ldots,A_n, A \in\{ \Bool, \Nat, \State\}$. We say that $\mathcal{R}$ is a {\em functional set of reduction rules} for $C$ if $\mathcal{R}$ consists, for all $c\in C$ and all ${a_1}:A_1,\ldots, {a_n}:A_n$ closed normal terms of $\SystemT_\State$, of exactly one rule $c {a_1}\ldots {a_n}\mapsto {a}$, for some closed normal term ${a}:A$ of $\SystemT_\State$.
\end{defi}

\begin{thm} \label{theorem-ExtStrongNormalization}
Assume that $\mathcal{R}$ is a functional set of reduction rules for $C$ (def. \ref{definition-functional}). Then $\SystemTState + C + \mathcal{R}$ enjoys strong normalization and weak-Church-Rosser (uniqueness of normal forms) for all closed terms of atomic types.

\end{thm}

\proof \emph{(Sketch)} For strong normalization, see \cite{Berger} (the constants ${s}:\State$ and $c \in C$ are trivially strongly computable). For weak Church-Rosser property, we start from the fact that there is the canonical set-theoretical model $\mathcal{M}$ of $\SystemTState + C + \mathcal{R}$. The interpretation of $\Bool, \Nat,\State$ in $\mathcal{M}$ consists of all closed normal form of these types. Arrows and pairs are interpreted set-theoretically. Each constant $c \in C$ is interpreted by some map $f_c$, defined by $f_c(a_1,\ldots,a_n)=a$ for all reduction rules $(c a_1\ldots a_n\mapsto a) \in \mathcal{R}$. Assume $u,v:A$ are closed normal term, $A = \Bool,\Nat$, or $\State$ is an atomic type, and $u,v$ are equal in $\SystemTState + C + \mathcal{R}$, in order to prove that $u, v$ are the same term. $u,v$ are equal in $\mathcal{M}$ because $\mathcal{M}$ is a model of $\SystemTState + C + \mathcal{R}$. By induction on $w$ we prove that if $w$ is a closed normal form of atomic type $\SystemT + C + \mathcal{R}$, then $w$ is a numeral, or $\True,\False$, or a state constant, and therefore $w$ is interpreted by itself in $\mathcal{M}$. From $u,v$ equal in $\mathcal{M}$ we conclude that $u,v$ are the same term of $\SystemTState + C + \mathcal{R}$.
\qed

We define two extensions of $\SystemT_\State$: an extension $\SystemTClass$ with symbols denoting the non-computable maps $X_P, \Phi_P$ and no computable reduction rules, another extension $\SystemTLearn$, with the computable approximations $\chi_P,\phi_P$ of $X_P, \Phi_P$, and a computable set of reduction rules. We use the elements of $\SystemTClass$ to represent non-computable realizers, and the elements of $\SystemTLearn$ to represent a computable ``approximation'' of a realizer. In the next definition, we denote terms of type $\State$ by $\rho, \rho', \ldots$.

\begin{defi} \label{definition-TermLanguageL1}
Assume $P:\Nat^{k+1}\rightarrow \Bool$ is a $k+1$-ary predicate of $\SystemT$. We introduce the following constants:
\begin{enumerate}[(1)]

\item
$\chi_P:\State \rightarrow \Nat^k\rightarrow \Bool$
and
$\varphi_P:\State \rightarrow \Nat^k\rightarrow \Nat$.

\item
$X_P:\Nat^k\rightarrow \Bool$ and $\Phi_P: \Nat^k \rightarrow \Nat$.

\item
$\Cup:\State\rightarrow \State \rightarrow \State$.

\item
$\Add_P:\Nat^{k+1} \rightarrow \State$ and $\add_P:\State \rightarrow \Nat^{k+1} \rightarrow \State$.

\end{enumerate}

%
%

We denote $\Cup\rho_1\rho_2$ with $\rho_1\Cup\rho_2$. 
\begin{enumerate}[(1)]
\item
$\Xi_\State$ is the set of all constants $\chi_P,\varphi_P, \Cup, \add_P$.

\item
$\Xi$ is the set of all constants $X_P,\Phi_P, \Cup, \Add_P$.

\item
$\SystemTClass = \SystemT_\State + \Xi$.

\item
A term $t \in \SystemTClass$ has state $\makestate{\emptyset}$ if it has no state constant different from $\makestate{\emptyset}$.
\end{enumerate}
\end{defi}

Let $\vec{t} = t_1\ldots t_k$. We interpret $\chi_P{s} \vec{t}$ and $\varphi_P{s}\vec{t} $ respectively as a ``guess'' for the values of the oracle and the Skolem map $X_P$ and $\Phi_P$ for $\exists y.P\vec{t}y$, guess computed w.r.t. the knowledge state denoted by the constant $s$.  There is no set of computable reduction rules for the constants $\Phi_P, X_P \in \Xi$, and therefore no set of computable reduction rules for $\SystemTClass$. If $\rho_1, \rho_2$ denotes the states $S_1, S_2 \in \StateSet$, we  interpret $\rho_1 \Cup \rho_2$ as denoting the consistent union $S_1 \CupSem S_2$ of $S_1, S_2$. $\Add_P$ denotes the map constantly equal to the empty state $\emptyset$. $\add_P{\makestate{S}} \vec{n}m $ denotes the empty state $\emptyset$ if we cannot add the atom $\langle P, \vec{n},m\rangle$ to $S$, either because $\langle P,\vec{n},m'\rangle \in S$ for some numeral $m'$, or because $P\vec{n}m={\False}$. $\add_P{\makestate{S}} \vec{n}m $ denotes the state $\{\langle P, \vec{n},m \rangle\}$ otherwise. We define a system $\SystemTLearn$ with reduction rules over $\Xi_\State$ by a functional reduction set $\mathcal{R}_\State$.

\begin{defi} (The System $\SystemTLearn$) \label{definition-EquationalTheoryL1}
Let  $s, s_1, s_2$ be state constants denoting the states $S, S_1, S_2$. Let $\langle P, \vec{n},m \rangle$ be an atom. $\mathcal{R}_\State$ is the following functional set of reduction rules for $\Xi_\State$:
\begin{enumerate}[(1)]
\item
If $\langle P,\vec{n},{m}\rangle \in S$, then
$\chi_P{s}\vec{n} \mapsto {\True}$ and $\varphi_P{s}\vec{n} \mapsto {m}$, else
$\chi_P{s}\vec{n} \mapsto {\False}$ and $\varphi_P{s}\vec{n} \mapsto {0}$.


\item
${s_1}\Cup{s_2} \mapsto \makestate{S_1 \CupSem S_2}$
\item
$\add_P{s}\vec{n}{m} \mapsto \makestate{\emptyset}$ if either $\langle P,\vec{n},{m'} \rangle \in S$ for some numeral $m'$ or $P\vec{n}{m} = {\False}$, and $\add_P{s}\vec{n}{m} \mapsto \makestate{\{\langle P,\vec{n},{m} \rangle\}}$ otherwise.

\end{enumerate}

We define $\SystemTLearn = \SystemT_\State + \Xi_\State + \mathcal{R}_\State$.
\end{defi}

\textbf{Remark.} $\SystemTLearn$ is nothing but $\SystemTState$ with some ``syntactic sugar''. By Theorem \ref{theorem-ExtStrongNormalization}, $\SystemTLearn$ is strongly normalizing and has the weak Church-Rosser property for closed term of atomic types.  $\SystemTLearn$ satisfies a Normal Form Property.

\begin{lem}[Normal Form Property for $\SystemTLearn$] \label{lemma-normalform} Assume $A$ is either an atomic type or a product type. Then any closed normal term $t \in \SystemTLearn$ of type $A$ is: a numeral ${n}:\Nat$, or a boolean $\True,\False:\Bool$, or a state constant $s:\State$, or a pair $\langle u,v \rangle: B \times C$.
\end{lem}

\proof \emph{(Sketch)} By induction over $t$. For some $\vec{v}$, either $t$ is $(\lambda \vec{x}.u)(\vec{v})$, or $t$ is $\langle u,w\rangle(\vec{v})$, or $t$ is $x(\vec{v})$ for some variable $x$, or $t$ is $c(\vec{v})$ for some constant $c$, and either $c={0}, \mbox{S}, {\True}, {\False}, {s}, \linebreak R_T, {\tt if}_T, \pi_i$ is some constant of $\SystemTState$, or $c \in \Xi_\State$. If $t=(\lambda \vec{x}.u)(\vec{v})$, then $t$ has an arrow type if $\vec{v} = \emptyset$, while $t$ is not normal if $\vec{v} \not = \emptyset$. If $t=\langle u,w\rangle(\vec{v})$, then $\vec{v}=\emptyset$ and we are done.
If $t = x(\vec{v})$ then $t$ is not closed. The only case left is $t=c(\vec{u}):A$. $A$ is not an arrow type, therefore all arguments of $c$ are in $\vec{u}$. If $t=0$ we are done, if $t=\mbox{S}(u)$ we apply the induction hypothesis, if $t={\True},{\False}:\Bool$ or $t={s}: \State$ or $t = \langle u,v \rangle$ we are done. Otherwise either $t=R_T(n,f,a)\vec{t}, {\tt if}_T(b,a_1,a_2)\vec{t}, \pi_i(v)\vec{t}$, or $t = \chi_P(u,\vec{w}):\Nat$, or $t= \varphi_P(u,\vec{w}) :\Nat$, or $t=\Cup(u_1,u_2) :\State$, or $t=\add_P(u,\vec{w}): \State$. The proper subterms $n, w_1, \ldots, w_k:\Nat$, $b:\Bool$, $v:A \times B$, $u,u_1,u_2:\State$ of $t$ have atomic or product type and are closed normal. By induction hypothesis they are, respectively, a numeral, a boolean, a pair, a state constant. In all cases, $t$ is not normal. \qed

Let $t, t' \in \SystemTLearn$ be two closed terms of type $\State$. We abbreviate ``$t,t'$ denotes two states which are consistent and disjoint'' by: $t, t'$ are consistent and disjoint. $\makestate{\emptyset}, s$ are consistent and disjoint for every state constant $s$. The maps denoted by $\Cup, \add_P$ preserve the relation: ``to be consistent and disjoint''.

\begin{lem}\label{lemma-disjoint}
Assume $s, s_1, s_2$ are state constants and $\langle P, \vec{n},m\rangle$ is an atom.
\begin{enumerate}[\em(1)]
\item
$s, (\add_Ps\vec{n}m)$ are consistent and disjoint.
\item
Assume $s,s_1$ are consistent and disjoint, and $s,s_2$ are consistent and disjoint. Then $s, s_1 \Cup s_2$ are consistent and disjoint.
\end{enumerate}
\end{lem}

\proof\hfill
\begin{enumerate}[(1)]
\item
Assume $s$ denotes the state $S$. If $\add_Ps\vec{n}m$ denotes the empty state the thesis is immediate. Otherwise $\add_Ps\vec{n}m$ denotes $\{\langle P, \vec{n},m\rangle\}$ and $\langle P, \vec{n},m'\rangle \not \in S$ for all numeral $m'$. Then $\{\langle P, \vec{n},m\rangle\}$ is consistent and disjoint with $S$.
\item
By Lemma \ref{lemma-consistency}.\qed
\end{enumerate}

Each (in general, non-computable) term $t \in \SystemTClass$ is associated to a set $\{t[{s}]\ | $s$ \mbox{ is a } \linebreak \mbox{state constant}\} \subseteq \SystemTLearn$ of computable terms we call its ``approximations'', one for each state constant $s$.

\begin{defi} Assume $t \in \SystemTClass$ and  $s$ is a state constant. We call ``approximation of $t$ at state $s$'' the term $t[{s}]$ of $\SystemTLearn$ obtained from $t$ by replacing each constant $X_P$ with $\chi_P{s}$, each constant $\Phi_P$ with $\varphi_P{s}$, each constant $\Add_P$ with $\add_P{s}$.
\end{defi}

We interpret any $t[{s}] \in \SystemTLearn$ as a learning process evaluated w.r.t. the information taken from a state constant $s$ (the same $s$ for the whole term).
\\

Assume $t \in \SystemTClass$ is closed, $t:\State$ and $s$ is a state constant. Then $t[{s}]$ is a closed term of $\SystemTLearn$, and its normal form, by the Normal Form Property \ref{lemma-normalform}, is some state constant ${s}'$. We conclude $t[{s}]={s}'$ in $\SystemTLearn$. We prove that $s, s'$ are consistent and disjoint.

\begin{lem}\label{lemma-consistentdisjoint}
Assume $s$ is a state constant, $t \in \SystemTClass$,
$t:\State$ is closed, and all state constants in $t$ are consistent and disjoint with $s$.
\begin{enumerate}[\em(1)]
\item
If $t[s]$ reduces to $t'[s]$, then all state constants in $t'$ are consistent and disjoint with $s$.
\item
$s,t[s]$ are consistent and disjoint.
\item
If all state constants in $u$ are $\makestate{\emptyset}$, then $s, u[s]$ are consistent and disjoint.
\end{enumerate}
\end{lem}

\proof\hfill
\begin{enumerate}[(1)]
\item
It is enough to consider a one-step reduction. Suppose that $t[s]$ reduces to $t'[s]$ by contraction of a redex $r$ of $t[s]$. If $r$ is $(\lambda x u)t$ or $R_TuvS(w)$ or $ {\tt if}_T(b,a_1,a_2)$ or $\pi_i\langle v_1,v_2\rangle$ or $ \chi_Ps\vec{n}$, or $ \varphi_Ps\vec{n}$, then its contractum $r'$ does not contain any new state constant; hence, all state constants in $t'$ are consistent and disjoint with $s$. If $r$ is $s_1\Cup s_2$ or $\add_Ps\vec{n}m$, then both $s, s_1$ and $s,s_2$ are consistent and disjoint state constants by hypothesis on $t$; therefore, by Lemma \ref{lemma-disjoint}, in both cases $s$ and the contraction of $r$ are consistent and disjoint;  so all state constants in $t'$ are consistent and disjoint with $s$.
\item
Every reduct of $t[s]$ is $t'[s]$ for some $t' \in \SystemTClass$. If $t[s]$ reduces to a normal form $t'[s] \equiv s'$, then the only possibility is $t' \equiv s'$. By the previous point $1$, we conclude that $s'$ is consistent and disjoint with $s$.
\item
By the previous point $2$, and the fact that the only state constant $\makestate{\emptyset}$ in $u$ is consistent and disjoint with any $s$.\qed
\end{enumerate}

\noindent We introduce now a notion of convergence for families of
terms $\{t[{s_i}]\}_{i \in \NatSet} \subseteq \SystemTLearn$, defined
by some $t \in \SystemTClass$ and indexed over a set of state
constants $\{s_i\}_{i\in\NatSet}$. Informally, ``$t$ convergent" means
that $t[{s}]$ eventually stops changing when the knowledge state $s$
increases. If $s, s'$ are state constants denoting $S, S' \in
\StateSet$, we write $s \le s'$ for $S \subseteq S'$. We say that a
sequence $\{s_i\}_{i\in\NatSet}$ of state constants is a weakly
increasing chain of states (is w.i. for short), if $s_i\le s_{i+1}$
for all $i\in\NatSet$.

\begin{defi}
\label{definition-Convergence}(Convergence). Assume
that $\{s_i\}_{i\in\NatSet} $ is a w.i. sequence of state constants,
and $u, v \in \SystemTClass$.
\begin{enumerate}[(1)]

\item
  $u$ converges in $\{s_i\}_{i\in\NatSet}$ if $\exists i\in\NatSet.
\forall j\geq i.u[s_j]=u[s_{i}]$ in $\SystemTLearn$.

\item
$u$ converges if $u$ converges in every w.i. sequence of state constants.
\end{enumerate}
\end{defi}

Remark that if  $u$ is convergent, we do not ask that $u$
is convergent to the {\em same} value on {\em all} w.i. chain of states. The value learned by $u$ may depend on the information contained in the particular chain of state constants by which $u$ gets the knowledge. The chain of states, in turn, is selected by the particular definition we use for the ``learning strategy'' $\CupSem$. Different ``learning strategies'' may learn different values.

\begin{thm}[Stability Theorem] \label{theorem-StabilityTheorem}
Assume $t \in \SystemTClass$ is a closed term of atomic type $A$ ($A\in\{\Bool,\Nat,\State\}$). Then $t$ is convergent.
\end{thm}

\proof \emph{(Classical)}. Assume $S$ is any consistent and possibly infinite set of atoms. We define some (in general, \underline{not} computable) functional reduction set $\mathcal{R}(S)$ for the set $\Xi$ of constants and for $\SystemTClass$. The reductions for $X_P, \Phi_P, \Add_P$ are those for $\chi_P, \phi_P, \add_P$ in $\SystemTLearn$:

\begin{enumerate}[(1)]
\item
If $\langle P,\vec{n},{m}\rangle \in S$, then
$(X_P\vec{n}\mapsto {\True}), (\Phi_P\vec{n}\mapsto{m}) \in \mathcal{R}(S)$, else \linebreak
$(X_P\vec{n}\mapsto {\False}), (\Phi_P\vec{n}\mapsto {0}) \in \mathcal{R}(S)$


\item
$\Add_P\vec{n}{m} \mapsto \makestate{\emptyset}$ if either $\langle P,\vec{n},{m'} \rangle \in S$ for some numeral $m'$ or $P\vec{n}{m} = {\False}$, and $\Add_P\vec{n}{m} \mapsto \makestate{\{\langle P,\vec{n},{m} \rangle\}}$ o.w..
\end{enumerate}
and the reduction for $\Cup$ in $\mathcal{R}(S)$ is the reduction for $\Cup$ in $\SystemTLearn$. By theorem \ref{theorem-ExtStrongNormalization}, $\SystemTClass + \mathcal{R}(S)$ is strongly normalizing and weak-CR for all closed terms of atomic type, for any consistent set of atoms $S$. For the rest of the proof, let $\{s_i\}_{i\in\NatSet}$ be a w.i. chain of state constants. Assume $t \in \SystemTClass$ is a closed term of atomic type $A$. {\em Claim}. For any state constant $s$, the map $u \mapsto u[{s}]$ is a bijection from the reduction tree of $t$ in $\SystemTClass + \mathcal{R}(s)$ to the reduction tree of $t[{s}]$ in $\SystemTLearn$. {\em Proof of the Claim.} By induction over the reduction tree of $t[{s}]$. Every reduction $\beta,\pi,{\tt if}_T, R_T, \Cup$ over $t[{s}]$ may be obtained from the same reduction over $t$. All occurrences of $\chi_P, \varphi_P, \add_P$ in the reduction tree of $t[{s}]$ are of the form  $\chi_P{s}, \varphi_P{s}, \add_P{s}$, therefore every reduction over $\chi_P, \varphi_P, \add_P$ may be obtained from the corresponding reduction over $X_P,\Phi_P,\Add_P$.

Assume $a$ is the (unique, by weak-CR) normal form of $t$ in $\SystemTClass + \mathcal{R}(s)$. By the Claim, $a[{s}]$ is the normal form of $t[{s}]$ in $\SystemTLearn$. Since $a$ is normal in $\SystemTClass + \mathcal{R}(s)$, there is no $X_P,\Phi_P,\Add_P$ in $a$. Thus $a$ and $a[{s}]$ are the same term: $t$ and $t[{s}]$ have the same normal form respectively in $\SystemTClass + \mathcal{R}(s)$ and in $\SystemTLearn$. Let $\{s_i\}_{i \in \NatSet}$ be a given sequence of state constants. Define $S_\omega = \cup_{i \in \NatSet}S_i$, where $S_i$ is the state denoted by $s_i$. By strong normalization, the reduction tree of $t$ in $\SystemTClass + \mathcal{R}(S_\omega)$ is finite. Therefore in this reduction tree are used only \textit{finitely many} reduction rules from $\mathcal{R}(S_\omega)$, and for some numeral $n$ it is equal to the reduction tree of $t$ in $\SystemTClass + \mathcal{R}(s_n)$, and in $\SystemTClass + \mathcal{R}(s_m)$ for all $m \ge n$. We deduce that for all $m \ge n$ the normal forms of $t$ in $\SystemTClass + \mathcal{R}(s_m)$ are the same. Thus, the normal form in $\SystemTLearn$ of all $t[s_m]$ with $m \ge n$ are the same, as we wished to show.
\qed


\begin{rem} The idea of the proof of theorem \ref{theorem-StabilityTheorem} corresponds exactly to the intuition of the introduction.  During any computation, the oracles $X_P$ and $\Phi_P$ are consulted a finite number of times and hence asked for a finite number of values. When our state of knowledge is great enough, we can substitute the oracles with their approximation $\chi_P{s}$ and $\varphi_P{s}$ for some state constant $s$, and we will obtain the same oracle values and hence the same results.

The proof, though non constructive, is short and well explains why the result is true. However, provided we replace the notion of convergence used in this paper with the intuitionistic notion introduced in \cite{Ber2005},
we are able to reformulate and prove theorem \ref{theorem-StabilityTheorem} in a purely intuitionistic way, achieving thus a constructive description of learning in $\HA +\EM_1$. Being the intuitionistic proof way more elaborated and less intuitive than the present one and connected with other foundationally interesting results, it will be the subject of a next  paper.

Our proof of convergence follows the pattern
of Avigad's one in \cite{Avigad}.
A closed term $t\in \SystemTClass$ of atomic
type and in the constant $c_1,\ldots, c_n\in \Xi$,
may be seen as a functional $F_t$ which
maps functions $f_1,\ldots, f_n$ of the same type
of $c_1,\ldots c_n$  into an object of atomic type:
$F_t(f_1,\ldots, f_n)$ is defined as the normal
form of $t$ in $\SystemTClass + \mathcal{R}$,
where $\mathcal{R}
=\{ c_ia_1\ldots a_n\mapsto a\ |\ f_i(a_1,\ldots,a_n)=
a \mbox{ and $i\in\{1,\ldots, n\}$}\}$. $F_t$
is continuous in the sense of Avigad. Moreover, since $X_P$ and $\Add_P$ have a set-theoretical definition in terms of $\Phi_P$, we may assume $F_t$ depends only on the functions which define in $\mathcal{R}$ the reduction rules for $\Phi_{P_1},\ldots\Phi_{P_n}$. Then, if $t$ is
of type $\State$, it is not difficult to see that
$F_t$ represents an update procedure with respect to any of its argument.
The fact that $F_t$ is an update procedure implies convergence
for $t$ and the fixed point property of theorem
\ref{Fixed Point Property}.
\end{rem}

Assume that $s$ is a state constant and $t\in\SystemTClass$  any closed term of type $\State$ of state $\makestate{\emptyset}$ (i.e., without state constants different from $\makestate{\emptyset}$). Denote by $\tau$ the map $:\StateSet \rightarrow \StateSet$ interpreting $s \mapsto t[s]$. $\tau$ is defined by $\tau(S) = S'$ if and only if $t[\makestate{S}] = \makestate{S}'$ in $\SystemTLearn$. By Lemma \ref{lemma-consistentdisjoint}, $S, \tau(S)$ are consistent and disjoint. In particular, $f(S) = S \cup \tau(S)$ defines a map $\fix:\StateSet \rightarrow \StateSet$. By Theorem \ref{theorem-StabilityTheorem}, if $\{S_i\}_{i \in \Nat}$ is any w.i. sequence of states, then
$\exists i. \forall j \ge i. \tau(S_j) = \tau(S_i)$.

As last result of this section, we prove that if we start from any state $S$, and we repeatedly apply $\fix:\StateSet \rightarrow \StateSet$, eventually we reach a state $S'=\fix^n(S)$ such that $\fix(S') = S'$ and $\tau(S)=\emptyset$. We interpret this result by saying that $\fix$ is a ``learning process'' adding the knowledge computed by the map $\tau$, and $\fix$  eventually stops extending the knowledge.

\begin{thm}[Fixed Point Property]\label{Fixed Point Property}
Let $t:\State$ be a closed term of $\SystemTClass$ of state $\makestate{\emptyset}$, and $s = \makestate{S}$. Define $\tau(S) = S'$ if $t[\makestate{S}] = \makestate{S}'$, and $f(S) = S \cup \tau(S)$.
\begin{enumerate}[\em(1)]
\item
There are $h\in\NatSet$, $S'\in\StateSet$ such that ${S}' = \fix^h({S})\supseteq S$, $\fix({S}')={S}'$ and $\tau(S') = \emptyset$.
\item
We may effectively find a state constant $s' \ge s$ such that $t[s'] = \makestate{\emptyset}$.
\end{enumerate}
\end{thm}

\proof\hfill
\begin{enumerate}[(1)]
\item
$\fix^0({S}),\fix^1({S}),
\fix^2({S}),\ldots$ is a w.i. chain of states because $f(S') \supseteq S'$ for all $S' \in \StateSet$. By theorem \ref{theorem-StabilityTheorem}, the map $\tau:\StateSet \rightarrow \StateSet$, interpreting the map $s \mapsto t[s]$, converges over this chain: there exists $k\in\NatSet$ such that for every $j\geq k$, $\tau(\fix^j({S}))=\tau(\fix^k({s}))$. By definition of $f$ and the choice of $k$:
\[\fix^{k+2}({S})
=
\fix^{k+1}({S}) \cup \tau(\fix^{k+1}({S}))
=
(\fix^{k}({S})
\cup
\tau(\fix^k({S})))
\cup \tau(\fix^k({S}))
=\]
\[=\fix^{k}({S})
\cup
\tau(\fix^k({S}))
=
\fix^{k+1}({S})
\]
Choose ${S}' = \fix^{k+1}({S})$. By the line above, we have $S'  \ge S$ and $\fix({S}') = {S}'$, therefore $\tau(S') \subseteq \fix(S') = S'$. From $S', \tau(S')$ disjoint we conclude $\tau(S')=\emptyset$.
\item
By the previous point and $t[\makestate{S}'] = \makestate{\emptyset}$ if and only if $\tau(S')=\emptyset$.\qed

\end{enumerate}

\section{An Interactive Learning-Based Notion of
Realizability}\label{section-ALearningBasedRealizability}
In this section we introduce the notion of realizability for $\HA +
\EM_1$, Heyting Arithmetic plus Excluded Middle on
$\Sigma^0_1$-formulas, then we prove our Main Theorem, the Adequacy
Theorem: {\em ``if a closed arithmetical formula is provable in
$\HA
+ \EM_1$, then it is realizable''}. \IfPaperState{For proofs we
refer to
\cite{ExtendedVersion}.}{}

We first define the formal system $\HA + \EM_1$, from now on ``Extended $\EM_1$ Arithmetic''. We represent atomic predicates of $\HA + \EM_1$ with (in general, non-computable) closed terms of $\SystemTClass$ of type $\Bool$. Terms of $\HA + \EM_1$ may include function symbols $X_P$, $\Phi_P$ denoting non-computable functions: oracles and Skolem maps for $\Sigma^0_1$-formulas $\exists x.Px\vec{{n}}$, with $P$ predicate of $\SystemT$. We assume having in $\SystemT$ some terms $\Rightarrow_\Bool: \Bool,\Bool\rightarrow\Bool, \neg_\Bool: \Bool \rightarrow \Bool, \ldots$, implementing boolean connectives. If $t_1, \ldots, t_n, t \in \SystemT$ have type $\Bool$ and are made from free variables all of type $\Bool$, using boolean connectives, we say that $t$ is a tautological consequence of $t_1, \ldots, t_n$ in $\SystemT$ (a tautology if $n=0$) if all boolean assignments making $t_1, \ldots, t_n$ equal to ${\True}$ in $\SystemT$ also make $t$ equal to ${\True}$ in
$\SystemT$.

\begin{defi} \label{definition-extendedarithmetic} (Extended $\EM_1$ Intuitionistic Arithmetic: $\HA + \EM_1$)
The language $\LanguageClass$ of $\HA + \EM_1$ is defined as follows.
\begin{enumerate}[(1)]

\item
The terms of $\LanguageClass$ are all $t \in \SystemTClass$ with state $\makestate{\emptyset}$, such that $t:\Nat$ and $FV(t) \subseteq \{x_1^\Nat, \ldots, x_n^\Nat\}$ for some $x_1, \ldots, x_n$.

\item
The atomic formulas of $\LanguageClass$ are all $Qt_1\ldots t_n \in \SystemTClass$, for some $Q:\Nat^{n}\rightarrow \Bool$ {\em closed term of $\SystemTClass$} of state $\makestate{\emptyset}$, and some terms $t_1,\ldots,t_n$ of $\LanguageClass$.

\item
The formulas of $\LanguageClass$ are built from atomic formulas of $\LanguageClass$ by the connectives $\lor,\land,\rightarrow \forall,\exists$ as usual.
\end{enumerate}


A formula of $\HA$ is a formula of $\HA + \EM_1$ in which all predicates and terms are terms of $\SystemT$.

Deduction rules for $\HA + \EM_1$
are as in van Dalen \cite{van Dalen}, with: 
{\em (i)} an axiom schema for $\EM_1$;
{\em (ii)} the induction rule;
{\em (iii
)} as Post rules:
all axioms of equality and ordering on $\Nat$, all
equational axioms of $\SystemT$, and one schema for each
tautological consequences of $\SystemT$. {\em (iv)} the axiom schemas for oracles: $P(\vec{t},t) \Rightarrow_\Bool X_P \vec{t}$ and for Skolem maps: $X_P \vec{t} \Rightarrow_\Bool P( \vec{t},(\Phi_P \vec{t}))$, for any predicate $P$ of $\SystemT$.

\end{defi}

We denote with $\bot$ the atomic formula ${\False}$ and will sometimes write  a generic atomic formula as $P(t_1,\ldots, t_n)$ rather than in the form $Pt_1\ldots t_n$. Finally, since any arithmetical formula has only variables of type $\Nat$, we shall freely omit their types, writing for instance $\forall x. A$ in place of $\forall x^\Nat. A$. Post rules cover many rules with atomic assumptions and conclusion as we find useful, for example, the rule: ``if $f(z)\leq 0$ then $f(z)=0$''.


We defined $\Rightarrow_\Bool:\Bool,\Bool\rightarrow \Bool$ as a term implementing implication, therefore, to be accurate, the axiom $P(t_1,\ldots, t_n,t) \Rightarrow_\Bool X_P t_1\ldots t_n$ is not an implication between two atomic formulas, but it is equal to the single atomic formula $Qt_1\ldots t_nt$, where \[ Q = \lambda x_1^{\Nat}\ldots \lambda
x_{n+1}^{\Nat}. \Rightarrow_\Bool (Px_1\ldots x_nx_{n+1})(X_P x_1\ldots x_{n+1})\]
Similarly, $\lnot_{\Bool}P(\vec{t},t)$ will denote a single atomic formula. Any atomic formula $A$ of $\LanguageClass$ is a boolean term of $\SystemTClass$, therefore for any state constant $s$ we may form the ``finite approximation'' $A[{s}]:\Bool, A[{s}] \in \SystemTLearn$ of $A$. In $A[{s}]$ we replace all oracles $X_P$ and all Skolem maps $\Phi_P$ we have in $A$ by their finite approximation $\chi_P{s}, \phi_Ps$, computed with respect to the state constant $s$. We denote with $\LanguageLearn$ the set of all expressions $A[s]$ with $A \in \LanguageClass$ and $s$ a state constant. All $A[s] \in \LanguageLearn$ may be interpreted by first order arithmetical formulas having all closed atomic subformulas decidable.

Using the metaphor explained in the introduction, we use a set of falsifiable hypotheses determined by $s$ to predict a computable truth value $A[{s}]:\Bool$ in $\SystemTLearn$ for an atomic formula $A \in \LanguageClass$ that we cannot effectively evaluate. Our definition of realizability  provides a formal semantics  for the Extended
Intuitionistic Arithmetic $\HA+\EM_1$, and therefore also for the more usual language of Arithmetic $\HA$, in which all functions represent recursive maps.

\begin{defi}
\label{definition-TypesForRealizers} (Types for realizers) For each
arithmetical formula $A$ we define a type $|A|$ of $\SystemT$ by
induction on $A$:
$|P(t_1,\ldots,t_n)|=\State$,
$|A\wedge B|=|A|\times |B|$,
$|A\vee B|= \Bool\times (|A|\times |B|)$,
$|A\rightarrow B|=|A|\rightarrow |B|$,
$|\forall x A|=\Nat\rightarrow |A|$,
$|\exists x A|= \Nat\times |A|$
\end{defi}

We define the realizability relation $t\Vvdash A$, where $t \in \SystemTClass$, $A \in \LanguageClass$, $t$ has state $\makestate{\emptyset}$ and $t:|A|$. The realizer denotes a non-computable map $t$, and is associated to a family $\{t[{s}]|s \mbox{ state constant}\}$ of one computable map $t[{s}]$ for each $s$, realizing the approximation $A[s] \in \LanguageLearn$ of the formula $A$. We interpret the set of Excluded Middle instances and Skolem axioms effectively used by a given proof as a set of experiments checking the assumptions we have in $s$ about Skolem maps and oracles. If all experiments succeed, the realizer provides a ``construction'' for $A$; if some experiment fails, the realizer provides some new knowledge obtained from the failure.


We first define $t\Vdash_s A$, the realizability relation  for the ``approximations'' $t \in \SystemTLearn$ and $A \in \LanguageLearn$, w.r.t. any state constant $s$, then we define $t'\Vvdash A'$ for $t' \in \SystemTClass$ of state $\makestate{\emptyset}$ and any closed $A' \in \LanguageClass$. For any types $T,U,V$, let $\proj_0, \proj_1, \proj_2$ denote the three projections from $T \times (U \times V)$.

\begin{defi}
\label{definition-AuxiliaryRealizability} Let $s$ be the constant denoting a state $S \in \StateSet$. Assume $t \in \SystemTLearn$ and $A \in \SystemTLearn$ are of the form $t = t'[s], A = A'[s]$ for some closed $t' \in \SystemTClass$ of state $\makestate{\emptyset}$ and some closed $A' \in \LanguageClass$. We define $t \Vdash_s A$ for any state constant $s$ by induction on $A$.

\begin{enumerate}[(1)]
\item
$t\Vdash_s P(t_1,\ldots, t_n)$ if and only if  $t = \makestate{\emptyset}$ in $\SystemTLearn$ implies
$P(t_1,\ldots,t_n)={\True}$

\item
$t\Vdash_s {A\wedge B}$ if and only if $\pi_0t\Vdash_s{A}$ and
$\pi_1t\Vdash_s{B}$

\item
$t\Vdash_s {A\vee B}$ if and only if: either $p_0t={\True}$ in $\SystemTLearn$ and
$p_1t\Vdash_s{A}$, or $p_0t={\False}$ and $p_2t\Vdash_s{B}$

\item
$t\Vdash_s {A\rightarrow B}$ if and only if for all $u$, if $u\Vdash_s{A}$,
then $tu\Vdash_s{B}$

\item
$t\Vdash_s {\forall x A}$ if and only if for all numerals $n$,
$t{n}\Vdash_s A[{n}/x]$

\item
$t\Vdash_s {\exists x A}$ if and only if  for some numeral $n$ $\pi_0t= {n}$ in $\SystemTLearn$ and $\pi_1t\Vdash_s A[{n}/x]$

\end{enumerate}
Assume $t' \in \SystemTClass$ is a closed term of state $\makestate{\emptyset}$, $A'\in \LanguageClass $ is a closed formula, and $t':|A'|$. We define
\begin{enumerate}[(1)]
\item
$t'\Vvdash_s A'$ if and only if $t'[{s}]\Vdash_s A'[{s}]$
\item
$t' \Vvdash A'$ if and only if $t'\Vvdash_s A'$ for all state constants $s$.
\end{enumerate}
\end{defi}

The realizability relation is compatible with equality in $\SystemTLearn$:

\begin{lem}\label{RealizabilityEquality}

If $t_1\Vdash_s A[u_1/x]$, $t_1=t_2$ and $u_1=u_2$ in $\SystemTLearn$, then $t_2\Vdash_s A[u_2/x]$

\end{lem}

\proof By straightforward induction on $A$.
\qed

By unfolding the definition of $t\Vvdash_s A$, we may obtain a direct characterization of the realizability relation for terms $t$ of $\SystemTClass$, bypassing the reference to the relation $\Vdash_s$ over ``approximations'' of terms and formulas of $\LanguageClass$. The only clause for $t\Vvdash_s A$ which is (slightly) different from the clause for $t\Vdash_s A$ is the clause for atomic formulas. We write the characterization of $\Vvdash$ explicitly because we refer to it in the next discussion.

%

\begin{lem}[Realizability]
\label{lemma-IndexedRealizabilityAndRealizability}
Assume $s$ is a state constant, $t\in \SystemTClass$ is a closed term, $A \in \LanguageClass $ is a closed formula, and $t:|A|$. Let $\vec{t} = t_1, \ldots, t_n : \Nat$.

\begin{enumerate}[\em(1)]
\item
$t\Vvdash_s P(\vec{t})$ if and only if $t[s]  = \makestate{\emptyset}$ in $\SystemTLearn$ implies
$P(\vec{t})[{s}]={\True}$

\item
$t\Vvdash_s{A\wedge B}$ if and only if $\pi_0t \Vvdash_s{A}$ and $\pi_1t\Vvdash_s{B}$

\item
$t\Vvdash_s {A\vee B}$  if and only if either $\proj_0t[{s}]={\True}$ in $\SystemTLearn$ and $\proj_1t\Vvdash_s A$, or $\proj_0t[{s}]={\False}$ in $\SystemTLearn$ and $\proj_2t\Vvdash_s B$

\item
$t\Vvdash_s {A\rightarrow B}$ if and only if for all $u$, if $u\Vvdash_s{A}$,
then $tu\Vvdash_s{B}$

\item
$t\Vvdash_s {\forall x A}$ if and only if for all numerals $n$,
$t{n}\Vvdash_s A[{n}/x]$
\item

$t\Vvdash_s \exists x A$ if and only for some numeral $n$, $\pi_0t[{s}]= {n}$ in $\SystemTLearn$ and $\pi_1t \Vvdash_s A[{n}/x][s]$

\end{enumerate}
\end{lem}

\proof By definition unfolding.
\qed

The characterizations of $\Vvdash$ shows that the definition of $\Vvdash$ formalizes all the idea we sketched in the introduction. A realizer is a term $t$ of $\SystemTClass$, possibly containing the non-computable functions $X_P, \Phi_P$; if such functions were computable, $t$ would be an intuitionistic realizer. Since in general $t$ is not computable, we calculate its approximation $t[s]$ at state $s$, which is a term of $\SystemTLearn$, and we require it to satisfy the indexed-by-state realizability clauses. Realizers of disjunctions and existential statements provide a witness, which is an
individual depending on an actual state of knowledge, representing all the hypotheses
used to approximate the non-computable. The actual behavior
of a realizer depends upon the current state of knowledge. The state is used only when there is relevant information about the truth of a given formula to be computed: the truth value $P(t_1,\ldots,t_n)[{s}]$ of an atomic formula and the disjunctive  witness $\proj_0t[{s}]$ and the existential witness $\pi_0u [{s}]$ are computed w.r.t. the constant state $s$. A realizer $t$ of $A\lor B$ uses $s$ to predict which one between $A$ and $B$ is realizable (if $\proj_0t[{s}]={\True}$ then $A$ is realizable, and if $\proj_0t[{s}]={\False}$ then $B$ is realizable). A realizer $u$ of $\exists  x A$ uses $s$ to predict that $\pi_0u[{s}]$ equals an ${n}$, some witness for $\exists x A$ (i.e. that $A[{n}/x]$ is realizable). These predictions need not  be always correct; hence, it is possible that a realized atomic formula is actually false; we may have $t \Vvdash_s P$ and $P[s]={\False}$ in $\SystemTLearn$. If an atomic formula, although predicted to be true, is indeed false, then we have encountered a counterexample and so our theory is wrong, our approximation  still inadequate; in this case, $t[{s}] \not = \makestate{\emptyset}$ by definition of $t\Vvdash_s P$, and the atomic realizer $t$ takes $s$ and extends it to a larger state $s'$, union of $s$ and $t[{s}]$. That is to say: if something goes wrong, we must learn from our
mistakes. The point is that after every learning, the actual state of knowledge grows, and if we ask to the same realizer new predictions, we will obtain ``better'' answers.

Indeed, we can say more about this last point. Suppose for instance that $t\Vvdash A\lor B$ and let $\{s_i\}_{i\in\NatSet}$ be a w.i. sequence. Then, since $t\in |\Bool|\times |A|\times |B|$, then $\proj_0t:\Bool$ is a closed term of $\SystemTClass$, converging in $\{s_i\}_{i\in\NatSet}$ to a boolean; thus the sequence of predictions  $\proj_0t[s_n]$ eventually stabilizes, and hence a witness is eventually learned in the limit.

In the atomic case, in order to have $t\Vvdash_s P(t_1,\ldots, t_n)$, we require that if $t[{s}] = \makestate{\emptyset}$, then $P(t_1,\ldots,t_n)[{s}] = {\True}$ in $\SystemTLearn$. That is to say: if $t$ has no new information to add to $s$,  then $t$ must assure the truth of $P(t_1,\ldots,t_n)$ w.r.t. $s$. By the Fixed Point Property (theorem \ref{Fixed Point Property}), when $t:\State$ is closed, there is plenty of state constants $s$ such that $t[s]=\makestate{\emptyset}$; hence search for truth will be for us {\em computation of a fixed point}, driven by the excluded-middle instances and the Skolem axioms used by the proof, rather than exhaustive search for counterexamples.\\

\begin{exa} The most remarkable feature of our Realizability Semantics is the existence of a $E_P$ realizer for $\EM_1$. Assume that $P$ is a predicate of $\SystemT$ and define $E_P$ as \[\lambda \vec{\alpha}^{\Nat}
\langle  X_P\vec{\alpha},\ \langle
\Phi_P\vec{\alpha},\
\makestate{\emptyset} \rangle ,\ \lambda n^{\Nat}\
\Add_P \vec{\alpha}n\rangle \rangle  \]

\begin{prop}\label{RealizerOfEM1}
(Realizer $E_P$ of $\EM_1$) $E_P\Vvdash \forall \vec{x}.\ \exists y\ P(\vec{x}, y)\vee \forall y \neg_\Bool P(\vec{x},y).$
\end{prop}

\proof Let $\vec{m}$ be a vector of numerals and let $s = \makestate{S}$ be a state constant denoting $S \in \StateSet$. $E_P\vec{m} [{s}]$ is equal to \[ \langle  \chi_P{s}\vec{m},\ \langle \varphi_P{s}\vec{m},\ \makestate{\emptyset} \rangle ,\ \lambda n^{\Nat}\ \add_P{s}\vec{m}n \rangle  \] and we want to prove that \[E_P\vec{m}[s]\Vdash_s  \exists y\ P(\vec{m}, y)\vee \forall y \neg_\Bool P(\vec{m},y)\]
We have $\proj_0E_P\vec{m}[{s}]=\chi_P{s}\vec{m}$ in $\SystemTLearn$. Assume $\chi_P{s} \vec{m}={\True}$. Then $\langle P, \vec{m}, {n} \rangle \in S$ for some numeral $n$ such that $P(\vec{m},{n}) = {\True}$, and we have to prove \[\proj_1 E_P{m}[{s}] \Vdash_s  \exists y\ P(\vec{m}, y)\] By definition unfolding, $\proj_1E_P{m}[{s}] = \langle \varphi_P{s} \vec{m},\makestate{\emptyset}\rangle =$ (by definition of $\varphi_P(s,\vec{m})$) $ \langle {n}, \makestate{\emptyset} \rangle$, hence, $\pi_0(\proj_1E_P{m})[{s}] =  \pi_0(\langle {n}, \makestate{\emptyset} \rangle) = {n}$ and $\proj_1(\pi_1E_P{m})[s] \Vdash_s P(\vec{m}, {n})$ because $P(\vec{m}, {n})={\True}$. We conclude $\proj_1E_P{m}[{s}] \Vdash_s  \exists y\ P(\vec{m}, y)$. Now assume $\proj_0E_P\vec{m} [{s}] = \chi_P{s} \vec{m}={\False}$. Then $\langle  P, \vec{m},n'\rangle  \not
\in S$ for all numerals $n'$.
We have to prove
\[\proj_2E_P\vec{m}[{s}]=\lambda n\ \add_P{s}\vec{m}n\
\Vdash_s
\forall y \neg_\Bool P({m},y)\] that is that, given any numeral $n$, \[ \add_ P{s}\vec{m}{n}  \Vdash_s \neg_\Bool
P({m},{n})\] By the definition of realizer in this case, we have to assume that $\add_P{s} \vec{m}{n} = \makestate{\emptyset}$, in order to prove that $\neg_\Bool
P(\vec{m}, {n}) [{s}] = {\True}$. The substitution $(.)[{s}]$ has an empty effect over $P(\vec{m}, {n})$, therefore we have to prove that $\neg_\Bool P(\vec{m}, {n}) = {\True}$, that is, that $P(\vec{m},{n})= {\False}$. Assume for contradiction that $P(\vec{m},{n}) ={\True}$. We already proved that $\langle P, \vec{m}, {n}' \rangle \not \in S$, for all numeral $n'$: from this and $P(\vec{m},{n}) ={\True}$ we deduce $\add_P {s} \vec{m}{n} = \makestate{\{\langle P, \vec{m}, {n} \rangle \}}$, contradiction.
\qed

%
%

$E_P$ works according to the ideas we sketched in the introduction.
It uses $\chi_P$ to make predictions about which one between
$\exists y\ P(\vec{m}, y)$ and $\forall y \neg_\Bool P(\vec{m},y)$ is true. $\chi_P$, in turn, relies on  the constant $s$ denoting the actual state to make its own prediction. If $\chi_P{s}{m}={\False}$, given any $n$, $ \neg_\Bool P({m},{n})$ is predicted to be true; if it is not
the case, we have a counterexample and $\Add_P$ requires to extend the state with $\langle P,\vec{m},{n}\rangle $. On the contrary, if $\chi_P{s}{m}={\True}$, there is unquestionable evidence that $\exists y P(\vec{m},y)$ holds; namely, there is some numeral $n$ such that $\langle P,\vec{m},{n}\rangle$ is in $s$; then $\varphi_P$ is called, and it returns $\varphi_P{s} \vec{m}={n}$.

This is the basic mechanism by which we implement learning: every state extension is linked with an assumption about an
instance of $\EM_1$ which we used and turned out to be wrong (this is the only way to come across a counterexample); in next computations, the actual state will be bigger, the realizer will not do the same error,  and hence will be ``wiser".
\end{exa}

\begin{exa} ($\Pi_2^0$ \textbf{formulas}) As usual for a Realizability interpretation, we may extract from any
realizer $t\Vvdash {\forall x. \exists y. P(x,y)}$, with $P \in \SystemT$, some recursive map $\psi$ from the set of numerals to the set of numerals, such that $P(n,\psi(n))$ for all numerals $n$. Indeed, by unfolding the definition of realizer, for all numerals $n$, all state constants $s$,
 $\pi_1(tn)[{s}]\Vdash_s
P(n,\pi_0(tn)[{s}])$. $\pi_1(tn)$ has state $\makestate{\emptyset}$ because $t$ is a realizer. Let us define $\tau(S) = S'$ if and only if $\pi_1(tn)[\makestate{S}] = \makestate{S}'$, and $\fix(S) = S \cup \tau(S)$, as in the proof of the Fixed
Point Theorem. Set $\phi(n) = \fix^k(\emptyset)$ for the first $k \in \NatSet$ such that $\fix^{k+1}(\emptyset) = \fix^{k}(\emptyset)$. Then $\phi(n)=\fix(\phi(n)) = \phi(n) \cup \tau(\phi(n))$, and by $\phi(n), \tau(\phi(n))$ disjoint we deduce $\tau(\phi(n))=\emptyset$, that is, $\pi_1(tn)[\makestate{\phi(n)}] = \makestate{\emptyset}$. By definition of realizer we have $P(n,\pi_0(tn)[\makestate{\phi(n)}]) = {\True}$  in $\SystemTLearn$. The required map $\psi$ is then
defined by $\psi(n) = \pi_0(tn)[\makestate{\phi(n)}]$ for all numerals $n$. We may prove that the map $\psi$ is definable in $\SystemTLearn$, and even in $\SystemT$, provided we replace the notion of convergence used in this paper with the intuitionistic notion of convergence introduced in \cite{Ber2005}, and we use this latter to provide a bound for the first $k \in \NatSet$ such that $\fix^{k+1} ({S}) = \fix^{k} ({S})$. We postpone this topic to another paper.
\end{exa}

\begin{rem} From the low level computational point of view and in the language of $\epsilon$-substitution method, our realizers represent convergent procedures to find out a ``solving substitution", i.e. a state representing an approximation of Skolem functions (i.e., $\epsilon$-terms) which makes true the Skolem axioms instances used in a proof of an existential statement. The advantage of our semantics is the possibility of defining such procedures directly from high level proofs, by means of Curry-Howard correspondence, hence avoiding the roundabout route which forces to use a quantifier free deduction system. In the case of a provable formula in the language of Peano Arithmetic (that is, one not containing the symbols $X_P$ or $\Phi_P$) we do not need at all to modify the language of its proof and to use the Skolem axioms $\chi, \varphi$.
\end{rem}

Now we explain how to turn each proof $\mathcal{D}$ of a formula $A
\in \LanguageClass$ in $\HA + \EM_1$ into a realizers $\mathcal{D}^*$
of the same $A$. By induction on $\mathcal{D}$, we define a
``decoration with
realizers'' $\mathcal{D}^\Realizer$ of $\mathcal{D}$, in which each
formula $B$ of $\mathcal{D}$ is replaced by a new statement $u
\vdash B$, for some $u \in \SystemTClass$ of state $\makestate{\emptyset}$. If $t \vdash A$ is the
conclusion of $\mathcal{D}^\Realizer$, we set $\mathcal{D}^* = t$.
Then we will prove that if $\mathcal{D}$ is closed and without
assumptions, then $\mathcal{D}^* \in \SystemTClass$ and
$\mathcal{D}^* \Vvdash A$. The decoration  $\mathcal{D}^\Realizer$
of
$\mathcal{D}$ with realizers is completely standard: we have new
realizers only for $\EM_1$ and for atomic formulas. For
notation simplicity, if $x_i$ is the label for the set of
occurrences of some assumption $A_i$ of $\mathcal{D}$, we use $x_i$
also as a name of one free variable in $\mathcal{D}^*$ of type
$|A_i|$. If $T$ is any type of $\SystemTState$, we denote with $d^T$ a dummy term of type $T$, defined by $d^\Nat = 0$, $d^\Bool = \False$, $d^\State = \makestate{\emptyset}$, $d^{A \rightarrow B} = \lambda \_^A.d^B$ (with $\_^A$ any variable of type $A$), $d^{A \times B} = \langle d^A, d^B\rangle$.

\begin{defi}
\label{definition-NaturalDeductionandTermAssignmentRulesforExtendedArithmetic}
(Term Assignment Rules for $\HA + \EM_1$).
Assume $\mathcal{D}$ is a proof of $A \in \LanguageClass$ in $\HA +
\EM_1$, with free assumptions $A_1, \ldots, A_n$ denoted by proof variables
$x_1^{A_1},
\ldots, x_n^{A_n}$ and free integer variables $\alpha_1^\Nat, \ldots,
\alpha_m^\Nat$. By induction on $\mathcal{D}$, we define a
decorated
proof-tree $\mathcal{D}^\Realizer$, in which each formula $B$
is replaced by $u \vdash B$ for some $u \in \SystemTClass$, and the
conclusion $A$ with some $t \vdash A$, with $FV(t) \subseteq
\{x_1^{|A_1|}, \ldots, x_1^{|A_1|},\alpha_1^\Nat, \ldots,
\alpha_m^\Nat \}$. Eventually we set $\mathcal{D}^*=t$.\\
%


%

\begin{enumerate}[(1)]

\item 
$\begin{array}{c} \mathstrut\\ \hline  x^{|A|}\vdash A\end{array}$
\hfill\break

\noindent if $\mathcal{D}$ consists of a single free
assumption $A\in \LanguageClass$ labeled $x^A$.\bigskip

\item 
$\begin{array}{c}  u\vdash A\ \ \ t\vdash B\\ \hline \langle
u,t\rangle \vdash
A\wedge B
\end{array}$\qquad
$\begin{array}{c}  u\vdash A\wedge B\\ \hline \pi_0 u\vdash A
\end{array}$\qquad
$\begin{array}{c}  u\vdash A\wedge B\\ \hline \pi_1 u\vdash B
\end{array}$\hfill\bigskip

\item 
$\begin{array}{c}  u\vdash A\rightarrow B\ \ \ t\vdash A \\ \hline
ut\vdash B
\end{array}$\qquad
$\begin{array}{c}  u\vdash B\\ \hline \lambda x^{|A|}u\vdash
A\rightarrow B
\end{array}$\hfill\bigskip

\item 
$\begin{array}{c}  u\vdash A\\ \hline \langle {\True},u,d^{B}\rangle
\vdash A\vee B
\end{array}$\qquad
$\begin{array}{c}  u\vdash B\\ \hline \langle  {\False},d^{A}, u\rangle
\vdash A\vee B
\end{array}$\hfill\break

\noindent$\begin{array}{c}  u\vdash A\vee B\ \ \ w_1\vdash C\ \ \ w_2\vdash
C\\ \hline  if\ \proj_0 u\ then\ (\lambda x^{|A|} w_1)(\proj_1u)\ else\
(\lambda x^{|B|} w_2)(\proj_2 u)\vdash C
\end{array}$\hfill\break

\noindent where $d^A$ and $d^B$ are dummy closed terms of
$\SystemTClass$ of type $|A|$ and $|B|$.\bigskip

\item 
$\begin{array}{c}  u\vdash \forall \alpha A\\ \hline  ut\vdash A[t/\alpha]
\end{array} $
$\begin{array}{c}  u\vdash A\\ \hline \lambda \alpha^{\Nat} u\vdash
\forall \alpha A
\end{array}$\hfill\break

\noindent where $t$ is a term of $\LanguageClass$ and $\alpha^{\Nat}$
does not occur free in any free assumption $B$ of the subproof of
$\mathcal{D}$ of conclusion $A$.\bigskip

\item 
$\begin{array}{c}  u\vdash A[t/\alpha^\Nat]\\ \hline  \langle
t,u\rangle \vdash
\exists
\alpha^\Nat. A
\end{array}$\qquad
$\begin{array}{c}  u\vdash \exists \alpha^\Nat. A\ \ \ t\vdash C\\
\hline
(\lambda \alpha^{\Nat}\lambda x^{|A|}\ t)(\pi_0 u)(\pi_1 u)\vdash C
\end{array} $\hfill\break

\noindent where $\alpha^{\Nat}$ is not free in $C$
nor in any free assumption $B$ different from $A$ in the
subproof of $\mathcal{D}$ of conclusion $C$.\bigskip

\item 
$\begin{array}{c}  u\vdash A(0)\ \ \ v\vdash \forall \alpha.
A(\alpha)\rightarrow A(S(\alpha))\\ \hline \lambda \alpha^{\Nat} Ruv\alpha\vdash
\forall
\alpha A
\end{array}$\hfill\bigskip

\item 
$\begin{array}{c}  u_1\vdash A_1\ u_2\vdash A_2\ \cdots \ u_n\vdash
A_n\\ \hline u_1\Cup u_2\Cup\cdots\Cup u_n\vdash A
\end{array}$\hfill\break

\noindent where $n > 0 $ and $A_1,A_2,\ldots,A_n,A$ are atomic
formulas of $\LanguageClass$, and the rule is a Post rule for equality
or ordering, or a tautological consequence.\bigskip

\item 
$\begin{array}{c} \mathstrut\\ \hline \makestate{\emptyset} \vdash A
\end{array}$\hfill\break

\noindent where $A$ is an atomic axiom of $\HA + \EM_1$ (an axiom
of equality or of ordering or a tautology or an equation of
$\SystemT$) \bigskip

\item 
$\begin{array}{c}\mathstrut\\\hline E_P \vdash  \forall \vec{x}.\ \exists y\
P(\vec{x}, y)\vee \forall y \neg_\Bool P(\vec{x},y)
\end{array}$\hfill\break

\noindent where $P$ is a predicate of $\SystemT$ and $E_P$ is defined
as $\lambda \vec{\alpha}^{\Nat} \langle X_P\vec{\alpha},\ \langle
\Phi_P\vec{\alpha},\ \makestate{\emptyset} \rangle ,\ \lambda
n^{\Nat}\ \Add_P \vec{\alpha}n\rangle \rangle$\bigskip

\item
$\begin{array}{c} \mathstrut\\\hline \Add_P \vec{t},t \vdash
P(\vec{t},t) \Rightarrow_\Bool X_P \vec{t}
\end{array}$\quad ($\chi$-Axiom)\bigskip

\item
$\begin{array}{c} \mathstrut\\\hline \makestate{\emptyset} \vdash X_P \vec{t}
\Rightarrow_\Bool P(\vec{t},(\Phi_P \vec{t}))
\end{array}$\quad ($\varphi$-Axiom)

\end{enumerate}
\end{defi}\bigskip

The term decorating the conclusion of a Post rule is of the form
$u_1\Cup \cdots \Cup u_n$. In this case, we have $n$ different
realizers, whose learning capabilities are put together through a
sort of union.
By Lemma \ref{lemma-Cup}.2, if $u_1\Cup \cdots \Cup u_n[{s}] = \makestate{\emptyset}$, then $u_1[{s}] = \ldots = u_n[{s}] = \makestate{\emptyset}$,
i.e. all $u_i$ ``have nothing to learn''. In that case, each $u_i$ must guarantee $A_i$ to be true, and therefore the conclusion of the Post rule is true, because true premises $A_1, \ldots, A_n$ spell a true conclusion
$A$.

We now prove our main theorem, that every theorem of $\HA +
\EM_1$ is realizable.

\begin{thm}[Adequacy Theorem]\label{Adeguacy Theorem}
Suppose that $\mathcal{D}$ is a proof of $A$ in
the system $\HA + \EM_1$ with free assumptions
$x_1^{A_1},\ldots,x_n^{A_n}$ and free variables
$\alpha_1:{\Nat},\ldots,\alpha_k:{\Nat}$. Let $w =
\mathcal{D}^*$.
For all state constants $s$ and for all numerals $n_1,\ldots,n_k$, if
\[t_1[s]\Vdash_s A_1[{n}_1/\alpha_1\cdots
{n}_k/\alpha_k][{s}], \ldots, t_n[s]\Vdash_s
A_n[{n}_1/\alpha_1\cdots
{n}_k/\alpha_k][{s}]\] then \[w[t_1/x_1^{|A_1|}\cdots
t_n/x_n^{|A_n|}\  {n}_1/\alpha_1\cdots
{n}_k/\alpha_k][{s}]\Vdash_s A[{n}_1/\alpha_1\cdots
{n}_k/\alpha_k][{s}]\]
\end{thm}

\IfPaperState
{
{\bf Proof} By induction on $w$ (see \cite{ExtendedVersion}).
}
{
\proof Notation: for any term $v$ and formula $B$, we
denote
\[v[t_1/x_1^{|A_1|}\cdots t_n/x_n^{|A_n|}\
{n}_1/\alpha_1\cdots
{n}_k/\alpha_k][{s}]\]
with $\substitution{v}$ and $B
[{n}_1/\alpha_1\cdots {n}_k/\alpha_k][{s}]$ with
$\substitution{B}$. We have $|\substitution{B}| = |B|$ for all formulas
$B$. We denote with $=$ the provable equality in $\SystemTLearn$.
We proceed by induction on $w$. Consider the last rule in the
derivation $\mathcal{D}$:

\begin{enumerate}[(1)]
\item
If it is the rule for variables, then
$w=x_i^{|A_i|}=x^{|\substitution{A_i}|}$ and
$A=A_i$. So $\substitution{w}=t_i\Vdash_s
\substitution{A_i}=\substitution{A}$.

\item
If it is the $\wedge I$ rule, then $w=\langle u,t\rangle $,
$A=B\wedge
C$, $u\vdash B$ and $t\vdash C$. Therefore, $\substitution{w}=
\langle \substitution{u},\substitution{t}\rangle $. By induction
hypothesis,
$\pi_0\substitution{w}=\substitution{u}\Vdash_s \substitution{B}$ and
$\pi_1\substitution{w}=\substitution{t}\Vdash_s \substitution{C}$; so, by
definition, $\substitution{w}\Vdash_s
\substitution{B}\wedge\substitution{C}=\substitution{A}$.

  \item If it is a $\wedge E$ rule, say left, then $w=\pi_0 u$ and
$u\vdash A\wedge B$. So $\substitution{w}=\pi_0 \substitution{u}\Vdash_s
\substitution{A}$, because $\substitution{u}\Vdash_s \substitution{A}\wedge
\substitution{B}$ by induction hypothesis.

  \item If it is the $\rightarrow E$ rule, then $w=ut$, $u\vdash
B\rightarrow A$ and $t\vdash B$. So
$\substitution{w}=\substitution{u}\substitution{t}\Vdash_s \substitution{A}$, for
$\substitution{u}\Vdash_s \substitution{B}\rightarrow \substitution{A}$ and
$\substitution{t}\Vdash_s \substitution{B}$ by induction hypothesis.

  \item If it is the $\rightarrow I$ rule, then $w=\lambda x^{|B|}
u$,
$A=B\rightarrow C$ and $u\vdash C$. Thus, $\substitution{w}=\lambda x^{|B|} \substitution{u}$. Suppose now that $t\Vdash_s
\substitution{B}$;
by induction hypothesis on $u$,
$\substitution{w}t=\substitution{u}[t/x^{|B|}] \Vdash_s \substitution{C}$.

%
%
%
%
%
%
%

\item If it is a $\vee I$ rule, say left, then $w=\langle
{\True},u,d^C\rangle $,
$A=B\vee C$ and $u\vdash B$. So,
$\substitution{w}=\langle {\True},\substitution{u},d^{C}\rangle $ and hence
$\proj_0\substitution{w}={\True}$. We indeed verify that
$\proj_1\substitution{w}=\substitution{u}\Vdash_s\substitution{B}$ with the help
of induction hypothesis.

\item If it is a $\vee E$ rule, then \[w= if\ \proj_0 u\ then\
(\lambda x^{|B|} w_1)\proj_1u\ else\ (\lambda y^{|C|} w_2)\proj_2u \]
 and  $u\vdash B\vee C, w_1\vdash D,w_2\vdash D, A=D$. So,
\[\substitution{w}= if\ \proj_0 \substitution{u}\ then\   (\lambda
x^{|B|}\substitution{w_1})(\proj_1\substitution{u})\ else\
(\lambda
y^{|C|}\substitution{w_2})(\proj_2 \substitution{u}) \]
Assume $\proj_0\substitution{u}={\True}$. Then by inductive hypothesis
$\proj_1 \substitution{u} \Vdash_s \substitution{B}$, and
again by induction hypothesis,
$\substitution{w}=\substitution{w}_1[\proj_1\substitution{u}/x^{|\substitution{B}|}]\Vdash_s
\substitution{D}$.
Symmetrically, if $\proj_0\substitution{u}={\False}$, then
$\substitution{w}\Vdash_s \substitution{D}$.

\item If it is the $\forall E$ rule, then $w=ut$, $A=B[t/\alpha]$
and $u\vdash \forall \alpha B$. So,
$\substitution{w}=\substitution{u}\substitution{t}$. For some numeral $n$ we have
${n}=\substitution{t}$. By inductive hypothesis  $\substitution{u}\Vdash_s
\forall\alpha \substitution{B}$, therefore
$\substitution{u}\substitution{t}=\substitution{u}{n}\Vdash_s
\substitution{B}[{n}/\alpha] = \substitution{B}[\substitution{t}/\alpha]=\substitution{A}$.

  \item If it is the $\forall I$ rule, then $w=\lambda
\alpha^{\Nat}u$, $A=\forall \alpha B$ and $u\vdash B$. So,
$\substitution{w}=\lambda \alpha^{\Nat} \substitution{u}$. Let $n$ be a numeral; we have to prove that
$\substitution{w}{n}=\substitution{u}[{n}/\alpha]\Vdash_s
\substitution{B}[{n}/\alpha]$, which is true, indeed, by
induction hypothesis.

  \item If it is the $\exists E$ rule, then $w=(\lambda
\alpha^{\Nat}\lambda x^{|B|} t)(\pi_0u)( \pi_1u)$, $t\vdash A$ and
$u\vdash \exists \alpha^{\Nat}. B$.

%
%

Assume ${n} = \pi_0
{u}$, for some numeral $n$. Then
\[\substitution{t}[{n}/\alpha^{\Nat},\pi_1\substitution{u}/
x^{|\substitution{B}[{n}/\alpha^{\Nat}]|}]\Vdash_s
\substitution{A}[{n}/\alpha]=A\] by
inductive hypothesis, whose application being justified by the
fact, also by induction, that $\substitution{u}\Vdash_s \exists
\alpha^{\Nat}.
\substitution{B}$ and hence $\pi_1\substitution{u}\Vdash_s
\substitution{B}[{n}/\alpha^{\Nat}]$. We thus obtain
\[\substitution{w}=\substitution{t}[\pi_0
\substitution{u}/\alpha^{\Nat}\
\pi_1\substitution{u}/x^{|B|}]
\Vdash_s
\substitution{A}[{n}/\alpha]=A\]

  \item If it is the $\exists I$ rule, then $w=\langle  t,u\rangle
$, $A=\exists
\alpha
B$, $u\vdash B[t/\alpha]$. So, $\substitution{w}=\langle
\substitution{t},\substitution{u}\rangle $; and, indeed, $\pi_1
\substitution{w}=\substitution{u}\Vdash_s \substitution{B}[\pi_0\substitution{w}/\alpha]=\substitution{B}[\substitution{t}/\alpha]$ since by induction hypothesis
$\substitution{u}\Vdash_s \substitution{B}[\substitution{t}/\alpha]$.

  \item If it is the induction rule, then $w=\lambda \alpha^{\Nat}\
Ruv\alpha$, $A=\forall \alpha B$, $u\vdash B(0)$ and $v\vdash
\forall \alpha. B(\alpha)\rightarrow B(S(\alpha))$. So,
$\substitution{w}=\lambda \alpha^{\Nat}
R\substitution{u}\substitution{v}\alpha$. Now let $n$ be a numeral. A
plain induction on $n$ shows that
$\substitution{w}{n}=R\substitution{u}\substitution{v}{n}\Vdash_s
\substitution{B}[{n}/\alpha]$, for $\substitution{u}\Vdash_s
\substitution{B}(0)$ and $\substitution{v}{i}\Vdash_s
\substitution{B}({i})\rightarrow
\substitution{B}({S(i)})$ for all numerals $i$ by
induction hypothesis.

  \item If it is a Post rule, then $w=u_1\Cup u_2\Cup \cdots\Cup
u_n$ and     $u_i\vdash  A_i$. So,
$\substitution{w}=\substitution{u}_1\Cup \substitution{u}_2\Cup \cdots\Cup
\substitution{u}_n$. Suppose now that $\substitution{w}[{s}] = \makestate{\emptyset}$; then we
have to prove that $\substitution{A}={\True}$. It suffices to prove that $\substitution{A}_1 =\substitution{A}_2 =\cdots= \substitution{A}_n ={\True}$. By Lemma \ref{lemma-Cup} we have $\substitution{u}_1 =\cdots= \substitution{u}_n =\makestate{\emptyset}$ and by induction hypothesis $\substitution{A}_1 =\cdots=\substitution{A}_n={\True}$, since $\substitution{u}_i\Vdash_s \substitution{A}_i$, for $i=1,\ldots,n$.

\item If it is a $\chi$-axiom rule, then  $w=\Add_Pt_1\ldots t_nt $ and \[A= P(t_1,\ldots, t_n,t)
\Rightarrow
X_P
 t_1\ldots t_n\] Let $\vec{t} = \substitution{t}_1,\ldots,\substitution{t}_n$. For some numeral $m$ we have $m = \substitution{t}$. Suppose by contradiction that $\substitution{w} = \makestate{\emptyset}$ and
$P(\vec{t},
\substitution{t}) = P(\vec{t},
{m}) ={\True}$ and $\chi_P s \vec{t}={\False}$. From $\chi_P s \vec{t}={\False}$ we get $\langle P,\vec{t}, {m}'\rangle \not \in s$ for all numerals $m'$. We deduce $\substitution{w} =\add_Ps\vec{t}
{m} = \makestate{\{\langle P, \vec{t},{m}\rangle\}}$, contradiction.

\item
$w$ realizes an $\EM_1$ axiom: this is Proposition \ref{RealizerOfEM1}.

\item If it is a $\varphi$-axiom rule, then $w=\makestate{\emptyset}$ and
\[A=X_P t_1\ldots t_n \Rightarrow P( t_1,\ldots,
t_n,(\Phi_P t_1\ldots t_n))\] We have
$\substitution{w}=\makestate{\emptyset}$. Let us denote $\vec{t} = {\substitution{t}_1}\ldots \substitution{t}_n$. Suppose that $\chi_P{s}\vec{t} ={\True}$. Then for some numeral $m$ we have $\langle P,\vec{t},{m}\rangle
\in s$ and $P\vec{t}{m} = {\True}$ and $\varphi_P{s}\vec{t} = {m}$. By definition of $\varphi_P$ we have  \[P( \vec{t},(\varphi_P{s} \vec{t}))={\True}\] We conclude that $\substitution{A} = {\True}$.\qed
\end{enumerate}
}


\begin{cor}\label{Realizability Theorem}
If $A$ is a closed formula provable in $\HA + \EM_1$, then there
exists $t\in \SystemTClass$ such that $t\Vvdash A$.
\end{cor}

\section{Conclusion and further works}
\label{section-conclusion}

Many notions of realizability for Classical Logic already exists. A notion similar to our one in spirit and motivations is Goodman's notion of Relative realizability \cite{Goodman}. However, there is an intrinsic difference between our solution and Goodman's solution. Goodman uses forcing to obtain a ``static'' description of learning. His ``possible worlds'' are learning states, but {\em there is no explicit operation updating a world to a larger word}. The dynamic aspect of learning (which is represented by a winning strategy in Game Semantics) is therefore lost. Using our realizability model, a realizer of an atomic formula, instead of being a trivial map, is {\em a map extending worlds}, whose fixed points are the worlds in which the atomic formula is true. Extending a world represents, in our realizability Semantics, the idea of ``learning by trial-and-error'' that we have in game semantics, while fixed points represent the final state of the game.

A second notion related to our realizability Semantics is Avigad's idea of ``update procedure'' \cite{Avigad}. A state $s$ in our paper corresponds to a finite model of skolem maps in Avigad. An ``update procedure'' is a construction ``steering'' the future evolution of a finite partial model $s$ of skolem maps, to which our individuals belong, in a wanted direction. The main difference with our work is that we express this idea formally, by interpreting an ``update procedure'' as a realizer (in the sense of Kreisel) for a Skolem axiom. Another important difference is that our realizability relation is defined for all first-order formulas with Skolem maps, while the theory of ``update procedures'' is defined only for quantifier-free formulas with Skolem maps.

Another difference with the other realizability or Kripke models for Classical Logic is in the notion of individual and in the equality between individuals. Assume that $m$ is the output of a skolem map for $\exists y.P(n,y)$, with $P$ decidable, and $m = \{m[s] | s \in \State\}$ a family of values depending on the finite partial model $s$. Then our realizer for Skolem axioms ``steers'' the evolution of $s$ towards some universe in which the axiom $\exists y.P(n,y) \Rightarrow P(n,m[s])$ is true. Modifying the evolution of $s$ may modify the value of $m[s]$. In our realizability Semantics we introduce a notion of individuality which is ``dynamical'' (depending on a state $s$) and ``interactive'' (the value of the individual depends on what a realizer does). This second aspect is new. A realizer may ``try'' to equate an individual $a = \{a[s] | s \in \State\}$ with another individual $b = \{b[s] | s \in \State\}$. Whenever this is possible, the realizer defines a construction over the evolution of the universe $s$ producing such an effect, while a random evolution of $s$ (without an ``interaction'' with the realizer) does not guarantee that eventually we have $a[s] = b[s]$. This is why, in our realizability model, even equality among concrete objects  is not a ``statical'' fact, but it is the effect of applying a realizer (which is a construction over the evolution of the state or ``world'' $s$). In the other models either equality is ``static'', or, even when it is ``dynamical'', and it changes with time, it is not ``interactive'': the final truth value of an equality is not the effect of the application of the realizer, but it is eventually the same in all future evolutions of the current world.
\\

Many aspects of our paper will require some further work. The first author is developing in his ph.d. thesis a constructive proof of the Fixed Point Property \ref{Fixed Point Property}, using the constructive notion of convergence introduced in \cite{Ber2005}. From a foundational viewpoint, this result will show that the sub-classical Arithmetic $\HA + \EM_1$ may be subsumed in Intuitionistic Arithmetic, in a sense.

Another challenging idea is to iterate the construction we had for $\EM_1$, in order to provide a learning model for the entire classical Arithmetic. In this case the leading concepts would be the game-theoretical notion of \emph{``level of backtracking''}, introduced in \cite{BerCoq} and \cite{BerardiLiguoro}, a notion related to the more informal notion of \emph{non-monotonic learning}.

Another aspect deserving further work is comparing the programs extracted from classical proofs with our method and with other methods, say, with Friedman $A$-translation. Our interpretation, explaining in term of learning how the extracted program work, should allow us to modify and improve the extracted program in a way impossible for the more formal (but very elegant) $A$-translation.

We remarked that our interpretation is implicitly parametric with respect to the operation $\CupSem$ merging the realizers of two atomic formulas. As explained in \cite{BerardiLiguoroMonadi}, by choosing different variant of this operation we may study different evaluation strategies for the extracted programs: sequential and parallel, left-to-right and right-to-left, confluent and non-confluent. We would like to study whether by choosing a particular evaluation strategy we may extract a more efficient program.


\begin{thebibliography}{99}


\bibitem{BerCoqKohlLICS}Y. Akama, S. Berardi, S.
Hayashi, U. Kohlenbach, \emph{An Arithmetical Hierarchy of the Law
of Excluded Middle and Related Principles}, in: LICS 2004, pp.
192-201.


\bibitem{AB} F. Aschieri, S. Berardi, \emph{Interactive Learning-Based Realizability Interpretation for Heyting Arithmetic with $\EM_1$},
Proceedings of TLCA 2009, Springer Lecture Notes in Computer
Science, vol. 5608, 2009



\bibitem{Ackermann}W. Ackermann,  \emph{Zur Widerspruchsfreiheit der Zahlentheorie},
Mathematische Annalen, 117, pp. 162Ð194 (1940)

\bibitem{Avigad} Jeremy Avigad: {\em Update Procedures and the
1-Consistency of Arithmetic}. Math. Log. Q. 48(1): 3-13 (2002).

\bibitem{Ber2005}S. Berardi,
\emph{Classical Logic as Limit Completion}, MSCS, Vol. 15, n.1,
2005, pp.167-200.

\bibitem{BerAPAL}S. Berardi,
\emph{Some intuitionistic equivalents of classical principles for
degree 2 formulas}, Annals of Pure and Applied Logic, Vol. 139,
n.1-3,
2006, pp.185-200.

\bibitem{BerCoq}S. Berardi, T. Coquand, S. Hayashi,
\emph{Games with 1-Bactracking},
APAL 2010, to appear. 

\bibitem{Berardi}S. Berardi, U. de' Liguoro,
\emph{A calculus of realizers for $\EM_1$-Arithmetic},
Proceedings of Computer Science Logic 2008,
in LNCS 5213, pag 215-229 (2008)

\bibitem{BerardiLiguoro} Stefano Berardi and Ugo de'Liguoro,
\emph{Toward the interpretation of non-constructive reasoning as non-monotonic learning}, Information and Computation, vol. 207, 1, pag. 63-81, (2009).

\bibitem{BerardiLiguoroMonadi} Stefano Berardi and Ugo de'Liguoro,
\emph{Interactive Realizers and Monads}, Draft, 2010. http://www.di.unito.it/~deligu/papers/InteractiveRealizers.pdf


\bibitem{Berger}U. Berger,
\emph{Continuous Semantics for Strong Normalization},
Lecture Notes in Computer Science 3526, 23--34, 2005

\bibitem{Coquand}T. Coquand,
\emph{A Semantic of Evidence for Classical Arithmetic},
Journal of Symbolic Logic 60, pag 325-337 (1995)

\bibitem{van Dalen} D. v. Dalen,
\emph{Logic and Structure},
Springer-Verlag, $3^{rd}$ Ed., Berlin Heidelberg (1994)

\bibitem{Girard}J.-Y. Girard,
\emph{Proofs and Types},
Cambridge University Press (1989)

\bibitem{Gold} E. M. Gold,
\emph{Limiting Recursion},
Journal of Symbolic Logic 30, pag. 28-48 (1965)
Cambridge University Press (1989)

\bibitem{Goodman} Nicolas D. Goodman,
\emph{Relativized Realizability in Intuitionistic Arithmetic of All Finite Types},Journal of Symbolic Logic 43, 1, pag. 23-44 (1978).


\bibitem{Hayashi0}S. Hayashi, R. Sumitomo, K.
Shii,
\emph{Towards Animation of Proofs -Testing Proofs by Examples - },
Theoretical Computer Science (2002)

\bibitem{Hayashi1}S. Hayashi,
\emph{Can Proofs be Animated by Games?},
FI 77(4), pag 331-343 (2007)

\bibitem{Hayashi}S. Hayashi,
\emph{Mathematics based on incremental learning - Excluded Middle
and Inductive Inference},
Theoretical Computer Science 350, pag 125-139  (2006)

\bibitem{Kleene}S. C. Kleene,
\emph{On the Interpretation of Intuitionistic Number Theory},
Journal of Symbolic Logic 10(4), pag 109-124 (1945)

\bibitem{Kreisel}G. Kreisel, \emph{Interpretation of analysis by means of constructive functionals of Þ-
nite types}, Heyting, A. (ed.), Constructivity in Mathematics, pp. 101Ð128. North-
Holland, Amsterdam (1959).

\bibitem{Popper}K. Popper,
\emph{The Logic of Scientific Discovery},
Routledge Classics, Routledge, London and New York (2002)
\end{thebibliography}
\end{document}